\documentclass[prd,preprint,superscriptaddress,amsmath,amssymb,nofootinbib]{revtex4}
\usepackage{graphicx,color,slashed}

\begin{document}

{\begin{flushright}{KIAS-P17016}
\end{flushright}}

\title{ Excesses of muon $g-2$, $R_{D^{(\ast)}}$, and $R_{K^{(*)}}$ in a leptoquark model}
%
%
\author{Chuan-Hung Chen}
\email{physchen@mail.ncku.edu.tw}
\affiliation{Department of Physics, National Cheng-Kung University, Tainan 70101, Taiwan}

\author{Takaaki Nomura}
\email{nomura@kias.re.kr}
\affiliation{School of Physics, KIAS, Seoul 02455, Korea}

\author{Hiroshi Okada}
\email{macokada3hiroshi@cts.nthu.edu.tw}
\affiliation{Physics Division, National Center for Theoretical Sciences, Hsinchu 300, Taiwan}

\date{\today}

\begin{abstract}
In this study, we investigate muon $g-2$, $R_{K^{(*)}}$, and $R_{D^{(*)}}$ anomalies in a specific model with one doublet, one triplet, and one singlet scalar leptoquark (LQ). When the strict limits from the $\ell' \to \ell \gamma$, $\Delta B=2$, $B_{s}\to \mu^+ \mu^-$, and $B^+ \to K^+ \nu \bar\nu$ processes are considered, it is difficult to use one scalar LQ to explain all of the anomalies due to the strong correlations among the constraints and observables. After ignoring the constraints and small couplings, the muon $g-2$ can be explained by the doublet LQ alone due to the $m_t$ enhancement, whereas the measured and unexpected smaller $R_{K^{(*)}}$ requires the combined effects of the doublet and triplet LQs, and the $R_D$ and $R_{D^*}$ excesses depend on the singlet LQ through scalar- and tensor-type interactions. 
  
\end{abstract}
\maketitle

\section{Introduction}

Several interesting excesses in semileptonic $B$ decays have been determined in experiments such as: (i) the angular observable $P'_5$ of $B\to K^* \mu^+ \mu^-$~\cite{DescotesGenon:2012zf}, where a $3.4\sigma$ deviation due to the integrated luminosity of 3.0 fb$^{-1}$ was found at the LHCb~\cite{Aaij:2015oid,Aaij:2013qta}, and the same measurement with a $2.6\sigma$ deviation was also reported by Belle~\cite{Wehle:2016yoi}; and (ii) the branching fraction ratios $R_{D,D^*}$, which are defined and measured as:
\begin{align}
R_{D}  = \frac{ \bar B\to D \tau \nu}{\bar B\to D \ell \nu} & = \left\{ 
\begin{tabular}{cc}
$0.375 \pm 0.064 \pm 0.026$ & \text{Belle~\cite{Huschle:2015rga} }\,,    \\
$0.440\pm 0.058\pm 0.042$ &  \quad ~ \text{BaBar~\cite{Lees:2012xj, Lees:2013uzd} }\,,
\end{tabular} \right. \nonumber \\
R_{D^*}  =  \frac{\bar B\to D^* \tau \nu}{\bar B\to D^* \ell \nu} & =\left\{ 
\begin{tabular}{cc}
$0.302\pm 0.030 \pm 0.011$ & \text{Belle~\cite{Abdesselam:2016cgx} }\,,    \\
$0.270 \pm 0.035 \pm^{+0.028}_{-0.025}$ & \text{Belle}~\cite{Hirose:2016wfn}\,, \\
$0.332 \pm 0.024 \pm 0.018$ &  \quad ~\text{BaBar~\cite{Lees:2012xj, Lees:2013uzd} }\,, \\
$0.336 \pm 0.027 \pm 0.030$ & \text{LHCb~\cite{Aaij:2015yra}}\,,
\end{tabular} \right.
\end{align}
where $\ell=(e,\, \mu)$, and these measurements can test the violation of lepton-flavor universality. 
The averaged results from the heavy flavor averaging group are $R_D=0.403\pm 0.040 \pm 0.024$ and $R_{D^*} =0.310\pm 0.015\pm 0.008$~\cite{Amhis:2016xyh}, and the standard model (SM) predictions are around $R_D \approx 0.3$~\cite{Lattice:2015rga,Na:2015kha} and $R_{D^*} \approx 0.25$, respectively. 

Further tests of lepton-flavor universality can be made using the branching fraction ratios $R_{K^{(*)}} = BR(B\to K^{(*)}\mu^+\mu^-)/BR(B\to K^{(*)} e^+e^-)$. The current LHCb measurements are $R_K = 0.745^{+0.090}_{-0.074}\pm 0.036$~\cite{Aaij:2014ora} and $R_{K^*} = 0.69^{+0.11}_{-0.07} \pm 0.05$~\cite{Aaij:2017vbb}, which indicate a more than $2.5\sigma$ deviation from the SM results. In addition, a known anomaly is the muon anomalous magnetic dipole moment (muon $g-2$), where its latest measurement is $\Delta a_\mu = a^{\rm exp}_\mu - a^{\rm SM}_\mu =(28.8\pm 8.0)\times 10^{-10}$~\cite{PDG}. If we assume that these results are correct, we need to extend the SM to explain these excesses. Inspired by these experimental observations, various solutions to the anomalies have been proposed~\cite{Fajfer:2012vx,Matias:2012xw,Descotes-Genon:2013wba,Sakaki:2013bfa,Gauld:2013qja,Datta:2013kja,Hurth:2013ssa,Duraisamy:2014sna,Descotes-Genon:2014uoa,Altmannshofer:2014rta,Descotes-Genon:2015uva,Crivellin:2015era,Lee:2015qra,Alonso:2015sja,Sahoo:2015qha,Sahoo:2015pzk,Chiang:2016qov,Dorsner:2016wpm,Boucenna:2016wpr,Hiller:2016kry,Crivellin:2015mga,Sahoo:2015wya,Straub:2015ica,Becirevic:2015asa,Hiller:2014yaa,Hurth:2014vma,Glashow:2014iga,Gripaios:2014tna,Sahoo:2015fla,Bauer:2015knc,Das:2016vkr,Li:2016vvp,Chen:2016dip,Becirevic:2016oho,Becirevic:2016yqi,Sahoo:2016pet,Bhattacharya:2016mcc,Duraisamy:2016gsd,Fajfer:2012jt,Crivellin:2012ye,Datta:2012qk,Deshpande:2012rr,Tanaka:2012nw,Ko:2012sv,Dorsner:2013tla,Abada:2013aba,Freytsis:2015qca,Crivellin:2015hha,Wei:2017ago,Cvetic:2017gkt,Ko:2017quv,Ko:2017yrd,Cheung:2016fjo,He:2012zp,Fajfer:2015ycq,Cheung:2017efc,Wang:2016ggf,Cheung:2016frv,Ivanov:2016qtw,Ivanov:2017mrj,Bardhan:2016uhr,ColuccioLeskow:2016dox}.

In the SM, the $b\to c \ell' \bar\nu_{\ell'}$ decays $(\ell'=e, \mu, \tau)$ arise from the $W$-mediated tree diagram, whereas the $b\to s \ell'^+ \ell'^-$ decays are generated by $W$-mediated box and $Z$-mediated penguin diagrams. 
 In the present study, based on our earlier study of muon $g-2$ and $R_K$ anomalies~\cite{Chen:2016dip}, we attempt to establish a specific model that simultaneously explains the muon $g-2$, $R_{K^{(*)}}$, and $R_{D^{(*)}}$ anomalies when the experimental bounds involved are satisfied. The serious constraints include $\ell_i \to \ell_j \gamma$, $\Delta F=2$, $B_s \to \mu^+ \mu^-$, $B\to K \nu \bar \nu$, etc. To clarify the effects introduced, we do not scan all of the parameters involved, but instead we retain the relevant couplings that can satisfy or escape from the experimental bounds, whereas we directly neglect the constrained and smaller couplings. 

To obtain the non-universal lepton-flavor effects, we consider the extension of the SM by including scalar leptoquarks (LQs), where the LQs are colored scalar particles that are coupled to a lepton and a quark at the same vertex, and the couplings to the quarks and leptons are flavor-dependent free parameters. LQs can couple to fermions and charge-conjugation of them at the same time, so in addition to the $SU(2)$ singlet, doublet, and triplet representations, the hypercharge of each representation may also have different choices depending on what quarks (leptons) or charge-conjugated quarks (leptons) couple to the LQs. Hence, in order to explain all of the excesses mentioned earlier in an actual model, we must decide what LQs are needed. 

It is known that the effective interactions for the muon $g-2$ can be expressed as $\bar \mu \sigma_{\alpha \beta} P_\chi \mu F^{\alpha\beta}$, where $P_\chi = P_{R(L)}$ is the chiral projection operator and $F^{\alpha\beta}$ is the electromagnetic field strength tensor. The initial and final muons carry different chirality, so in order to enhance the $\Delta a_\mu$ and avoid suppression by the lepton mass, the introduced LQ must interact with the left-handed and right-handed leptons. Due to the gauge invariance, the LQ can be an $SU(2)$ doublet, and its hypercharge can be determined as $Y=7/6$. 

 In addition to the muon $g-2$, the doublet LQ can also contribute to $b\to s \ell'^+ \ell'^-$; thus, this LQ may help resolve the excesses in $B\to K^{(*)} \mu^+ \mu^-$. Unfortunately, the corrections to the Wilson coefficients of $C_9$ and $C_{10}$ for the $b\to s \ell^+ \ell^-$ decays have the same sign, whereas we need an opposite sign to explain the measurements of the $R_{K^{(*)}}$, $P'_5$, and $B_s\to \mu^+ \mu^-$ decays. Moreover, when combined with the experimental limits, the Yukawa couplings involved are too small to explain the $R_D$ and $R_{D^*}$ anomalies. Thus, we have to introduce more LQs. 
 
 Due to the SM neutrinos being left-handed particles, the extra LQs for the $b\to c \ell' \bar\nu_{\ell'}$ processes must couple to the doublet leptons. According to the gauge invariance, these LQs can be singlet, doublet, or triplet. The $b\to c$ transition involves up- and down-type quarks, so the doublet
LQ is excluded as a candidate. A triplet LQ is a good candidate for the $b\to s \ell'^+ \ell'^-$ processes because the associated values for $C_9$ and $C_{10}$ have opposite signs. The triplet LQ can contribute to both $b\to s \ell'^+ \ell'^-$ and $b\to c \ell' \bar\nu_{\ell'}$ decays at the tree level, but it can be shown that both processes share the same LQ couplings. Therefore, by considering the constraints on the $b\to s \ell^+ \ell^-$ decays and $\Delta B=2$ process, the $R_{D}$ and $R_{D^*}$ cannot be enhanced significantly. Thus, in addition to the triplet LQ, it is necessary to consider a singlet LQ~\cite{Sakaki:2013bfa, Bauer:2015knc}. Intriguingly, we show that such a singlet LQ can contribute to $b\to c \ell' \bar\nu_{\ell'}$ but not to $b\to s \ell'^+ \ell'^-$ at the tree level, i.e., the couplings of the singlet LQ are not affected by the $b\to s \ell^+ \ell^-$ constraints.  
The singlet LQ can induce $b\to s \mu^+ \mu^-$ according to one-loop diagrams~\cite{Bauer:2015knc}, but a previous analysis by~\cite{Becirevic:2016yqi} showed that it is not a viable approach when using a singlet scalar LQ to simultaneously explain $R_{D^{(*)}}$ and $R_{K^{(*)}}$. Hence, more LQs are necessary to explain the anomalies. 

The remainder of this paper is organized as follows. In Section II, we introduce our model and derive formulae for the numerical analysis.
In Section III, we present the numerical analysis to show the parameter regions that correspond with anomalies in semileptonic $B$ decays. A summary is given in Section IV.

 \section{Model and formulae}

In this section, we begin by formulating the model, before studying the relevant phenomena of interest. The three LQs introduced are $\Phi_{7/6}=(2, 7/6)$, $\Delta_{1/3}=(3, 1/3)$, and $S^{1/3}=(1,1/3)$ under $(SU(2)_L, {U(1)_Y})$ SM gauge symmetry, where the doublet and triplet representations can be taken as:
\begin{align}
 \Phi_{7/6} = \begin{pmatrix} \phi^{5/3} \\ \phi^{2/3} \end{pmatrix}\,, \
\Delta_{1/3} = \begin{pmatrix} \delta^{1/3}/\sqrt{2} & \delta^{4/3} \\ \delta^{-2/3} & - \delta^{1/3}/\sqrt{2} \end{pmatrix}\,,
\end{align}
where the superscripts are the electric charges of the particles. Accordingly, the LQ Yukawa couplings to the SM fermions are expressed as: 
\begin{align}
\label{eq:lang_lq}
-L_{LQ} &=  \left[ \bar u \,{\bf V k} P_R \ell \phi^{5/3}  + \bar d\,  {\bf k} P_R \ell \phi^{2/3}  \right] + \left[ -\bar 
\ell\,   {\bf \tilde{k}}  P_R  u \phi^{-5/3} + \bar \nu \, {\bf \tilde{k}} P_R u \phi^{-2/3} \right]   \nonumber \\ 
& + \left[ \overline{u^c} \, {\bf V^*y}  P_L  \nu \delta^{-2/3} - \frac{1}{\sqrt{2}} \overline{ u^c}  \, {\bf V^* y}  P_L \ell \delta^{1/3} - \frac{1}{\sqrt{2}} \overline{d^c} \, {\bf  y} P_L \nu \delta^{1/3} - \overline{ d^c} \,  {\bf y} P_L  \ell \delta^{4/3} \right]  \,, \nonumber \\
& +  \left(  \overline{u^c} \, {\bf V^* \tilde y } P_L \ell - \overline{ d^c} \, {\bf \tilde{y}} P_L \nu + \overline{u^c} {\bf w} P_R \ell \right) S^{1/3} + 
h.c.\,,
\end{align}
where the flavor indices are suppressed, ${\bf V}\equiv U^u_L U^{d\dagger}_{L}$ denotes the Cabibbo--Kobayashi--Maskawa (CKM) matrix, $U^{u,d}_L$ are the unitary matrices used to diagonalize the quark mass matrices, and $U^d_L$ and $U^u_R$ have been absorbed into ${\bf k}$, ${\bf \tilde{k}}$, ${\bf  y}$, ${\bf \tilde{y}}$, and ${\bf w}$. In the model, we cannot generate the neutrino masses. Therefore, we treat the neutrinos as massless particles and their flavor mixing effects are rotated away. 
 There is no evidence for any new CP violation, so in the following, we treat the Yukawa couplings as real numbers.   

The scalar LQs can also couple to the SM Higgs via the scalar potential, and the cross section for the Higgs to diphoton can be modified in principle. However, the couplings of the LQs to the Higgs are different parameters and irrelevant to the flavors, so by taking proper values for the parameters, the signal strength parameter for the Higgs to diphoton can fit the LHC data. Hence, we do not discuss this issue in the present study, but a detailed analysis was given by~\cite{Chen:2016dip}. 

\subsection{Effective interactions for semileptonic $B$-decay}

 According to the interactions in Eq.~(\ref{eq:lang_lq}), we first derive the four-Fermi interactions for the $b\to c \ell' \bar\nu_{\ell'}$ and $b\to s \ell'^+ \ell'^-$ decays. For the $b\to c \ell' \bar\nu_{\ell'}$ processes, the induced current-current interactions from $k_{3j}\tilde{k}_{i2}$ and $\tilde{y}_{3i} w_{2j}$ are $(S-P)\times (S-P)$ and those from $y_{3i} y_{2j}$ and $\tilde{y}_{3i}\tilde{y}_{2j}$ are $(S-P)\times (S+P)$, where $S$ and $P$ denote the scalar and pseudoscalar currents, respectively. After taking the Fierz transformations, the Hamiltonian for the $b\to c \ell' \bar\nu_{\ell'}$ decays can be expressed as: 
\begin{align}
 {\cal H}_{b\to c}  &= \left( -\frac{ \tilde{y}_{3i} w_{2 j}}{2m^2_{S}}+\frac{ k_{3j} \tilde{k}_{i 2}}{2m^2_{\Phi}}  \right)  \bar c P_L b \, \bar\ell_j P_L \nu_i +  \left( \frac{ \tilde{y}_{3i} w_{2 j}}{2m^2_{S}}+\frac{ k_{3j} \tilde{k}_{i 2}}{2m^2_{\Phi}}  \right)  \frac{1}{4} \bar c \sigma_{\mu\nu} P_L b \, \bar \ell_j \sigma^{\mu \nu} P_L \nu_i  \nonumber \\
 & - \sum_a V_{2a} \frac{ y_{aj}y_{3i} }{4m^2_{\Delta}} \bar c \gamma_\mu P_L b\, \bar\ell_j \gamma^\mu P_L \nu_i + \sum_a V_{2a} \frac{\tilde{y}_{aj}\tilde{y}_{3i} }{2m^2_{S}} \bar c \gamma_\mu P_L b\, \bar\ell_j \gamma^\mu P_L \nu_i \,, \label{eq:Hbc}
 \end{align}   
where the indices $i,\,j$ are the lepton flavors, and the LQs in the same representation are taken as degenerate particles. It can be seen that the interaction structure obtained from the triplet LQ is the same as that from the $W$-boson. The doublet LQ generates an $(S-P)\times (S-P)$ structure, but also a tensor structure. However, the singlet LQ can produce $(V-A)\times (V-A)$, $(S-P)\times (S-P)$, and tensor structures. Nevertheless, we show later that the singlet LQ makes the main contribution to the $R_D$ and $R_{D^*}$ excesses. It is difficult to explain $R_{D, D^*}$ by only using the doublet or/and triplet LQs when the $R_K$ excess and other strict constraints are satisfied. 
 
 Using the Yukawa couplings in Eq.~(\ref{eq:lang_lq}), the effective Hamiltonian for the $b\to s \ell'^+ \ell'^-$ decays mediated by $\phi^{2/3}$ and $\delta^{4/3}$ at the tree level can be expressed as:
 \begin{align}
 {\cal H}_{b\to s} & = \frac{k_{3j} k_{2j}}{2 m^2_{\Phi}} (\bar{s}\gamma^\mu P_L b)(\bar{\ell}_j \gamma_\mu P_R \ell_j)\,, \nonumber \\
 &  -\frac{y_{3j} y_{2 j}}{2 m^2_{\Delta}} (\bar{s}\gamma^\mu P_L b)(\bar{\ell}_j \gamma_\mu P_L \ell_j)\,, \label{eq:H_bs}
 \end{align}
where the Fierz transformations have been applied. By Eq.~(\ref{eq:H_bs}), we can see clearly that the quark currents from both the doublet and triplet LQs are left-handed, whereas the lepton current from the doublet (triplet) LQ is right(left)-handed. When we include Eq.~(\ref{eq:H_bs}) in the SM contributions, the effective Hamiltonian for the $b\to s \ell'^+ \ell'^-$ decays is written as:
\begin{equation}
 {\cal H}_{b\to s }  = \frac{G_F \alpha_{\rm em} V_{tb} V^*_{ts}}{\sqrt{2}\pi} \left[ H_{1\mu} L^\mu + H_{2\mu} L^{5\mu} \right]\,, \label{eq:Hbsll}
 \end{equation}
where the leptonic currents are denoted by $L^{(5)}_\mu= \bar \ell \gamma_\mu (\gamma_5) \ell$, and the related hadronic currents are defined as:
 \begin{align}
 H_{1\mu} &= C^\ell_9 \bar s \gamma_\mu P_L b - \frac{2m_b}{q^2} C_7 \bar s i \sigma_{\mu \nu} q^\nu P_R b\,, \nonumber \\
 H_{2\mu} & = C^{\ell}_{10} \bar s \gamma_\mu P_L b\,.
 \end{align}
The effective Wilson coefficients with LQ contributions are expressed as:
 \begin{align}
 C^\ell_{9(10)} & = C^{\rm SM}_{9(10)} + C^{\rm LQ, \ell'}_{9(10)}\,,  \nonumber \\
  C^{LQ,\ell_j}_{9} &= -\frac{1}{4 c_{\rm SM}}\left( \frac{k_{3j} k_{2j}}{m^2_\Phi} - \frac{y_{3j} y_{2j}}{m^2_\Delta}\right)\,, \nonumber \\
  C^{LQ,\ell_j}_{10} & =- \frac{1}{4 c_{\rm SM}}\left( \frac{k_{3j} k_{2j}}{m^2_\Phi} + \frac{y_{3j} y_{2j}}{m^2_\Delta}\right)\,, \label{eq:Wilsons}
 \end{align}
where 
 $c_{\rm SM} = V_{tb} V^*_{ts} \alpha_{\rm em} G_F/(\sqrt{2} \pi)$, and $V_{ij}$ is the  CKM matrix element.
 From Eq.~(\ref{eq:Wilsons}), we can see that when the magnitude of $C^{LQ,\ell_j}_{10}$ is decreased, $C^{LQ,\ell_j}_{9}$ can be enhanced, i.e., the synchrony of the increasing/decreasing Wilson coefficients of $C^{\rm NP}_{9}$ and $C^{\rm NP}_{10}$ from new physics is diminished in this model. In addition, the sign of $C^{LQ,\ell'}_9$ can be different from that of $C^{\rm LQ,\ell'}_{10}$. Therefore, when the constraint from $B_s \to \mu^+ \mu^-$ decay is satisfied, we can obtain sizable values for $C^{LQ,\mu}_{9}$ to fit the anomalies of $R_{K^{(*)}}$ and angular observable in $B\to K^* \mu^+ \mu^-$. The LQs can contribute to the electromagnetic dipole operators, but since the effects occur through one-loop diagrams and they are also small, the associated Wilson coefficient $C_7$ comes mainly from the SM contributions.
 As mentioned earlier, the singlet LQ $S^{1/3}$ can also contribute to $C_{9,10}$ through box diagrams~\cite{Bauer:2015knc}. Using our notations, the results can be expressed as~\cite{Bauer:2015knc}:
 \begin{align}
%
  C^{{\rm box}, \ell}_{LL} & = \frac{m^2_t}{8\pi \alpha_{\rm em} m^2_S}|V_{3j} \tilde y_{j \ell }|^2 - \frac{\sqrt{2}}{64 C_{\rm SM} m^2_S } \left( \sum_i  \tilde y_{2i} \tilde y_{3i} \right) \sum_j |V_{jj'} \tilde y_{j' \ell }|^2 \,, \nonumber \\
  C^{{\rm box}, \ell}_{LR} & = \frac{m^2_t}{8\pi \alpha_{\rm em} m^2_S} \left( \ln\frac{m^2_S}{m^2_t} - f\left(\frac{m^2_t}{m^2_W}\right)\right) |w_{3 \ell }|^2 - \frac{\sqrt{2}}{64 C_{\rm SM} m^2_S } \left( \sum_i  \tilde y_{2i} \tilde y_{3i} \right) \sum_j |w_{ j \ell }|^2 \,, \label{eq:C910LQ}
 \end{align}
where the Wilsons coefficients $C^{\rm box,\ell}_{9,10}=(C^{{\rm box},\ell}_{LR} \pm C^{{\rm box},\ell}_{LL})/2$ and $f(x)=1 - 3/(x-1) + 3\ln(x)/(x-1)^2$. If we take $V_{3j} \tilde y_{j \ell} \sim V_{33} \tilde y_{3 \ell}$ and $\sum_j |V_{jj'} \tilde y_{j' \ell}|^2\sim |\tilde y_{2 \ell}|^2 + |\tilde y_{3 \ell}|^2$ by neglecting the small CKM matrix elements, then we obtain $C^{\rm box, \ell}_{LL} \approx 0.16 |\tilde y_{3\ell}|^2 -0.18 (|\tilde y_{2\ell}|^2 +|\tilde y_{3\ell}|^2)$ and $C^{\rm box, \ell}_{LR} \approx 0.46 |w_{3\ell}|^2 -0.18 \sum_j |w_{j\ell}|^2$, where $m_S\approx 1000$ GeV, $V_{ts}\approx -0.04$, and $\sum_i \tilde y_{2i} \tilde y_{3i} \sim 0.09$ from the $B^+ \to K^+ \nu \bar \nu$ constraint (see below) are used. 
To obtain $C^{\rm box,\mu}_9 \sim -1$, we need $\tilde y_{32},w_{32} \ll 1$ and $\tilde y_{22}, w_{22} \sim 5$, i.e., when we explain $R_{K^{(*)}}$ anomalies by using the $S^{1/3}$ LQ, the same effect will enhance the $b\to c \mu \bar\nu$ process such that $R_{D^{(*)}}$ is still lower than the experimental data. A detailed analysis was given by~\cite{Becirevic:2016oho}. In order to avoid enhancing the $b\to c \mu \bar \nu$ channel, we assume that the loop effects of Eq.~(\ref{eq:C910LQ}) are small, and the resolution to $R_{K^{(*)}}$ comes from other LQ contributions in Eq.~(\ref{eq:Wilsons}). 


\subsection{ Constraints from $\Delta F=2$, radiative lepton-flavor violating, $B ^+\to K^+ \nu \bar \nu$, $B_s \to \mu^+ \mu^-$, and $B_c \to \tau \nu$ processes }

Before we analyze the muon $g-2$, $R_{D^{(*)}}$, $R_{K^{(*)}}$ problems, we examine the possible constraints due to rare decays. First, we discuss the strict limits from the $\Delta F=2$ processes, such as $F-\bar F$ oscillation, where $F$ denotes the neutral pseudoscalar meson. Since $K-\bar K$, $D-\bar D$, and $B_d-\bar B_d$ mixings are involved, the first generation quarks and the anomalies mentioned earlier are associated with the second and third generation quarks. Therefore, we can avoid the constraints by assuming that $k_{1\ell'}\approx \tilde{k}_{\ell' 1} \approx y_{1\ell'} \approx \tilde{y}_{1\ell'} \approx w_{1i} \approx 0$ without affecting the analyses of $R_{D^{(*)}}$ and $R_{K^{(*)}}$. Thus, the relevant $\Delta F=2$ process is $B_s-\bar B_s$ mixing, where $\Delta m_{B_s} = 2|\langle \bar B_s  | {\cal H}| B_s \rangle| $ is induced from box diagrams and the LQ contributions can be formulated as: 
\begin{align}
\Delta m_{B_s}& \approx
\frac{C_{\rm box}}{(4\pi)^2}
\left[ \frac{5}{4} \left( \frac{\sum^3_{i=1}y_{3i} y_{2i} }{m_\Delta}\right)^2   + \left(\frac{\sum^3_{i=1} k_{3i} k_{2i}}{m_\Phi} \right)^2 \right] \nonumber \\
&+
\frac{C_{\rm box}}{(4\pi)^2}
\left[  \left( \frac{\sum^3_{i=1}\tilde{y}_{3i} \tilde{y}_{2i} }{m_S}\right)^2  
+ 2 \frac{\left( \sum^3_{i=1} y_{3i} \tilde y_{2i} \right)\left( \sum^3_{i=1}\tilde{y}_{3i} {y}_{2i} \right)
}{m_S^2-m_\Delta^2} 
\ln\left[\frac{m_S}{m_\Delta}\right] \right] \,,  
\end{align}  
where $C_{\rm box}= m_{B_s} f_{B_s}^2/3$, $f_{B_s}\approx 0.224$ GeV is the decay constant of $B_s$-meson~\cite{Dowdall:2013tga}, and the current measurement is $\Delta m^{\rm exp}_{B_s} = 1.17\times 10^{-11}$ GeV~\cite{PDG}. To satisfy the $R_{K^{(*)}}$ excess, the rough magnitude of LQ couplings is $|y_{3i} y_{2i}|\sim | k_{3i} k_{2i} |\sim 5\times 10^{-3}$. Using these parameter values, it can be shown that the resulting $\Delta m_{B_s}$ agree with the current data. However, $\Delta m_{B_s}$ can indeed constrain the parameters involved in the $b\to c \ell' \bar\nu_{\ell'}$ decays. Later, we discuss how this constraint can be satisfied. 


In addition to the muon $g-2$, the introduced LQs can also contribute to the lepton-flavor violating processes $\ell' \to \ell \gamma$, where the current upper bounds are $BR(\mu \to e \gamma) < 4.2\times10^{-13}$ and 
 $BR(\tau \to e (\mu) \gamma) <  3.3 (4.4)\times10^{-8}$~\cite{PDG}, and they can strictly constrain the LQ couplings. To understand the constraints due to the $\ell' \to \ell \gamma$ decays, we express their branching ratios (BRs) as:
 \begin{equation}
BR(\ell_b \to \ell_a \gamma) = \frac{48 \pi^3 \alpha_{\rm em} C_{ba} }{G_F^2 m_{\ell_b}^2} \left( \left| (a_R)_{ab} \right|^2 + \left| (a_L)_{ab} \right|^2 \right)
\end{equation}
with
 $C_{\mu e} \approx 1$, $C_{\tau e} \approx 0.1784$, and $C_{\tau \mu} \approx 0.1736$. 
 $(a_{R})_{ab}$ is written as: 
\begin{align}
(a_R)_{ab} \approx & \frac{3}{(4 \pi)^2} \int d[X]~  m_{t}\,  (F_{k \tilde k} -F_{ w \tilde{y} })_{ab} \,, \label{eq:aRL}
\end{align}
where $\int [dX] \equiv \int dx dy dz \delta(1-x-y-z)$, $(a_L)_{ab}$ can be obtained from $(a_R)_{ab}$ by using $(F_{\alpha \beta }^\dagger)_{ab}$ instead of $(F_{\alpha \beta})_{ab}$, and the function $F_{k \tilde k}$ is given by: 
\begin{align}
(F_{k \tilde k})_{ab}  &=   ({\bf V k})_{3 b} \tilde k_{a 3}  \left( \frac{5}{3} \frac{ x}{\Delta(m_{t}, m_\Phi)_{ab}} + \frac{2}{3} \frac{1-x}{\Delta(m_\Phi, m_{t})_{ab} }\right)\,, \nonumber \\
(F_{w \tilde{y}})_{ab}  &=   w_{3b} ({\bf V \tilde{y}})_{3 a} \left( \frac{1}{3} \frac{ x}{\Delta(m_{t}, m_S)_{ab}} + \frac{2}{3} \frac{1-x}{\Delta(m_S, m_{t})_{ab} }\right)\,, \nonumber \\
\Delta(m_1,m_2)_{ab}  &\approx   x m_1^2 +(y+z) m_2^2 \,. 
\end{align}
 We note that ${\bf V k}_{3b} \approx k_{3b}$ and ${\bf V \tilde{y}}_{3a} \approx \tilde{y}_{3a}$ due to $V_{ub, cb} \ll V_{tb} \approx 1$. From Eq.~(\ref{eq:lang_lq}), we can see that the doublet and singlet LQs can simultaneously couple to both left- and right-handed charged leptons, and the results are enhanced by $m_t$. Other LQ contributions are suppressed by $m_\ell$ due to the chirality flip in the external lepton legs, and thus they are ignored. Based on Eq.~(\ref{eq:aRL}), the muon $g-2$ can be obtained as:
 \begin{equation}
 \Delta a_\mu \simeq - {m_\mu} (a_L + a_R)_{a=b=\mu}\,. \label{eq:muong-2}
 \end{equation}
 
 
 As mentioned earlier, the singlet LQ does not contribute to $b \to s \ell'^+ \ell^- $ at the tree level, but it will induce the $b\to s \nu \bar \nu$ process, where the current upper bound is $B^+ \to K^{+} \nu \bar \nu < 1.6 \times 10^{-5}$, and the SM result is around $4\times 10^{-6}$. Therefore, $B^+\to K^+ \nu \bar\nu$ can bound the parameters of $\tilde{y}_{3i}\tilde{y}_{2i}$. The four-Fermi interaction structure, which is induced by the LQ, is the same as that induced by the $W$-boson, so we can formulate the BR for $B^+ \to K^+ \nu \bar\nu$ as:
 \begin{align}
& BR(B^+ \to K^+ \nu \bar \nu) \approx \frac{1}{3} \left(\sum_{\ell'} |1- r_{\ell'}|^2 \right) BR^{\rm SM}(B^+ \to K^+ \nu \bar \nu)\,, \\
& r_{\ell'} =  \frac{1}{C^\nu_{SM}}\left( \frac{\tilde{y}_{3\ell'} \tilde{y}_{2\ell'}}{2m^2_S} +  \frac{y_{3\ell'} y_{2\ell'}}{4m^2_\Delta} \right)\,, \quad C^{\nu}_{\rm SM} = \frac{G_F V_{tb} V^*_{ts} }{\sqrt{2}}  \frac{\alpha_{\rm em}}{2\pi \sin^2\theta_W}X(x_t) \,,
 \end{align}
where $x_t =m^2_t/m^2_W$ and $X(x_t)$ can be parameterized as $X(x_t) \approx 0.65 x^{0.575}_t$~\cite{Buchalla:1995vs}. According to Eq.~(\ref{eq:H_bs}), the LQs also contribute to $B_s \to \mu^+ \mu^-$, where the BRs measured by LHCb~\cite{Aaij:2017vad}
 and predicted by the SM~\cite{Bobeth:2013uxa} are $BR(B_s \to \mu^+ \mu^-)^{\rm exp}  = (3.0\pm 0.6^{+0.3}_{-0.2}) \times 10^{-9}$ and $BR(B_s \to \mu^+ \mu^-)^{\rm SM}  = (3.65 \pm 0.23) \times 10^{-9}$, respectively. The experimental data are consistent with the SM prediction, so in order to consider the constraint due to $B_s \to \mu^+ \mu^-$, we use the expression for the BR as~\cite{Hiller:2014yaa}:
 \begin{equation}
\frac{\text{BR}(B_s \to \mu^+ \mu^-)}{\text{BR}(B_s \to \mu^+ \mu^-)^{\text SM}} = \left|1-0.24 C^{LQ,\mu}_{10} \right|^2.
\end{equation}
 
 
 In addition to the $B^-\to D^{(*)} \tau \bar\nu$ decay, the induced effective Hamiltonian in Eq.~(\ref{eq:Hbc}) also contributes to the $B_c \to \tau \bar \nu$ process, where the allowed upper limit is $BR(B_c^- \to \tau \bar\nu ) < 30 \%$~\cite{Alonso:2016oyd}.
According to a previous given by~\cite{Alonso:2016oyd},  we express the BR for $B_c \to \tau \bar\nu$ as~\cite{Alonso:2016oyd}:
 \begin{align}
 BR(B_c \to \tau \bar \nu_\tau) = \tau_{B_c} \frac{m_{Bc} m^2_\tau f^2_{B_c} G_F^2 |V_{cb}|^2}{8 \pi} \left( 1 - \frac{m_\tau^2}{m_{B_c}^2} \right)^2 
 \left| 1 + \epsilon_L + \frac{m^2_{B_c}}{m_\tau (m_b + m_c)} \epsilon_P \right|^2,
 \label{eq:BRBc}
 \end{align}
 where $f_{B_c} $ is the $B_c$ decay constant and 
the $\epsilon_{L,P}$ in our model are given as:
 \begin{align}
 \epsilon_L &= \frac{\sqrt{2}}{4 G_F V_{cb}} \left[- \sum_a V_{2a} \frac{ y_{a3}y_{33} }{4m^2_{\Delta}} + \sum_a V_{2a} \frac{\tilde{y}_{a3}\tilde{y}_{33} }{2m^2_{S}} \right]\,, \nonumber \\
 \epsilon_P &= \frac{\sqrt{2}}{4 G_F V_{cb}} \left[ \frac{ \tilde{y}_{33} w_{2 3}}{2m^2_{S}} - \frac{ k_{33} \tilde{k}_{3 2}}{2m^2_{\Phi}} \right]\,. \nonumber 
 \end{align}
 Using $\tau_{B_c}\approx 0.507 \times 10^{-12}$s, $m_{B_c}\approx 6.275$ GeV, $f_{B_c}\approx 0.434$ GeV~\cite{Colquhoun:2015oha}, and $V_{cb}\approx 0.04$, the SM result is $BR^{\rm SM}(B_c \to \tau \bar \nu_\tau)\approx 2.1\%$. We can see that the effects of the new physics can enhance the $B_c\to \tau \bar \nu_\tau$ decay by a few factors at most in our analysis given in the following. 

 \subsection{ Observables: $R_{D^{(*)}}$ and $R_{K^{(*)}}$}

The observables of $R_{D^{(*)}}$ and $R_{K^{(*)}}$ are the branching fraction ratios that are insensitive to the hadronic effects, but the associated BRs still depend on the transition form factors. In order to calculate the BR for each semileptonic decay, we parameterize the transition form factors for $\bar B \to P$ as:
\begin{align} 
\langle P(p_2) | q \gamma^\mu b | \bar B(p_1) \rangle &= F_{+}(q^2) \left( (p_1 + p_2)^\mu -\frac{m^2_B -m^2_P}{q^2 } q^\mu \right)+ \frac{m^2_B - m^2_P}{q^2} q^\mu F_0(q^2)\,, \nonumber \\
 \langle P(p_2) | q \sigma_{\mu\nu} b | \bar B(p_1) \rangle &= - i (p_{1\mu} p_{2\nu} - p_{1\nu} p_{2\mu} ) \frac{2F_T(q^2) }{m_B + m_P}\,, \label{eq:ffP}
\end{align}
where $P$ can be the $D(q=c)$ or $K(q=s)$ meson, and the momentum transfer is given by $q=p_1 - p_2$. 
For the $B\to V$ decay where $V$ is a vector meson, the transition form factors associated with the weak currents are parameterized as:
 \begin{align}
 \langle V(p_2, \epsilon)| \bar q \gamma_\mu b|\bar B(p_1)\rangle & = i \varepsilon_{\mu \nu \rho \sigma} \epsilon^{\nu*} p^\rho_1 p^\sigma_2 \frac{2V(q^2) }{m_B + m_V}\,, \nonumber \\
 \langle V(p_2, \epsilon)| \bar q \gamma_\mu \gamma_5 b|\bar B(p_1)\rangle & = 2m_V A_{0}(q^2) \frac{\epsilon^*\cdot q}{q^2}q_\mu +
 (m_B + m_V)A_1(q^2) \left( \epsilon^*_\mu - \frac{\epsilon^* \cdot q}{q^2}q_\mu \right) \nonumber \\
 & - A_2(q^2) \frac{\epsilon^* \cdot q}{m_B + m_V} \left( (p_1 + p_2)_\mu - \frac{m^2_B -m^2_V}{q^2} q_\mu  \right)\,, \nonumber \\
\langle V(p_2, \epsilon)| \bar q \sigma_{\mu \nu} b|\bar B(p_1)\rangle & = \varepsilon_{\mu \nu \rho \sigma} \left[ 
 \epsilon^{\rho *} (p_1 + p_2)^\sigma T_1(q^2) + \epsilon^{\rho *} q^\sigma \frac{m^2_B -m^2_V}{q^2}(T_2(q^2) - T_1(q^2)) \right. \nonumber \\
 &\left. + 2 \frac{\epsilon^* \cdot q}{q^2} p^\rho_1 p^\sigma_2 \left( T_2 (q^2) - T_1(q^2) + \frac{q^2}{m^2_B -m^2_V} T_3(q^2) \right) \right]\,, \label{eq:ffV}
 \end{align}
where $V=D^*(K^*)$ when $q=c(s)$,  $\epsilon^{0123}=1$, $\sigma_{\mu \nu} \gamma_5 =( i/2 )\epsilon_{\mu \nu \rho \sigma} \sigma^{\rho \sigma}$, and $\epsilon^\mu$ is the polarization vector of the vector meson.  We note that the form factors associated with the weak scalar/pseudoscalar currents can be obtained through the equations of motion, i.e., $i\partial_\mu \bar q \gamma^\mu b = (m_b - m_q) \bar q b$ and $i\partial_\mu (\bar q \gamma^\mu \gamma_5 b) = - (m_b + m_q) \bar q \gamma_5 b$. For numerical estimations, the $q^2$-dependent form factors $F_+$, $F_T$, $V$, $A^0$, and $T_1$ are taken as~\cite{Melikhov:2000yu}:
 \begin{align}
 f(q^2) & = \frac{f(0)}{(1-q^2/M^2)(1-\sigma_1 q^2/M^2 + \sigma_2 q^4/M^4)} \,, \label{eq:FF1}
 \end{align}
and the other form factors are taken as:
  \begin{align}
 f(q^2) & = \frac{f(0)}{1-\sigma_1 q^2/M^2 + \sigma_2 q^4/M^4} \,. \label{eq:FF2}
 \end{align}
The values of $f(0)$, $\sigma_1$, and $\sigma_2$ for each form factor are shown in Table~\ref{tab:FF}. A detailed discussion of the form factors was given by~\cite{Melikhov:2000yu}. The NNL effects obtained with the LCQCD approach for the $B\to D$ form factors were described by~\cite{Wang:2017jow}.

\begin{table}[hpbt]
\caption{ $B\to P, V$ transition form factors, as parameterized in Eqs.~(\ref{eq:FF1}) and (\ref{eq:FF2}). }
\begin{ruledtabular}
\begin{tabular} {|c||ccc||ccccccc|} %
 & \multicolumn{3} {c||}{$B\to D$}  & \multicolumn{7} {c|}{$B\to D^*$}  \\ \hline
  &  $F_+$ & $F_0 $& $F_T$ &  $V$ & $A_0$ & $A_1$ & $A_2$ & $T_1$ & $T_2$ & $T_3$ \\ \hline
  f(0) & 0.67 & 0.67 & 0.69 & 0.76 & 0.69 & 0.66 & 0.62 & 0.68 & 0.68 & 0.33 \\  \hline
  $\sigma_1$ &   0.57 & 0.78 & 0.56 & 0.57 & 0.58 & 0.78 & 1.40 & 0.57  & 0.64 & 1.46    \\ \hline 
  $\sigma_2$ & & & & & & & 0.41 & & &  \\ \hline \hline
  &  \multicolumn{3} {c||}{$B\to K$}  & \multicolumn{7} {c|}{$B\to K^*$}  \\ \hline
            f(0)    & 0.36   & 0.36 & 0.35 & 0.44 & 0.45 & 0.36 & 0.32 & 0.39 & 0.39 & 0.27 \\  \hline
  $\sigma_1$ &   0.43 & 0.70 & 0.43 & 0.45 & 0.46 & 0.64 & 1.23 & 0.45  & 0.72 & 1.31    \\ \hline 
  $\sigma_2$ &          &   0.27 &        &         &         & 0.36 & 0.38 &          &  0.62 & 0.41 \\ 
\end{tabular}
\end{ruledtabular}

\label{tab:FF}
\end{table}

 According to the form factors in Eqs.~(\ref{eq:ffP}) and (\ref{eq:ffV}), and the interactions in Eqs.~(\ref{eq:Hbc}) and (\ref{eq:Hbsll}), we briefly summarize the differential decay rates for the semileptonic $B$ decays, which we use for estimating $R_{D^{(*)}}$ and $R_K$. For the $\bar B \to D \ell' \bar \nu_{\ell'}$ decay, the differential decay rate as a function of the invariant mass $q^2$ can be formulated as:
 \begin{align}
 \frac{d\Gamma^{\ell'}_D}{dq^2} & = \frac{G^2_F |V_{cb}|^2   \sqrt{\lambda_D}}{256 \pi^3 m^3_B} \left( 1- \frac{m^2_{\ell'}}{q^2}\right)^2 \left[ \frac{2}{3} \left( 2+ \frac{m^2_{\ell'}}{q^2}\right) |X^{\ell'}_+|^2 + \frac{2 m^2_{\ell'}}{q^2}  \left|X^{\ell'}_0 + \frac{\sqrt{q^2}}{m_{\ell'}} X^{\ell'}_S \right|^2 \right. \nonumber \\
 & \left.
 + 16  \left(\frac{2}{3}\left( 1+ \frac{2 m^2_{\ell'}}{q^2}\right) |X^{\ell'}_T|^2-\frac{m_{\ell'}}{\sqrt{q^2}} X^{\ell'}_T X^{\ell'}_0 \right) \right]\,, \label{eq:diffD}
 \end{align}
 where the $\{X^{\ell'}_\alpha \}$ functions and LQ contributions are given by:
 \begin{align}
 X^{\ell'}_+ & = \sqrt{\lambda_D}  (1 + C^{\ell'}_V) F_{+}(q^2) \,, \quad  X^{\ell'}_0 = (m^2_B -m^2_D)  (1+C^{\ell'}_V) F_0 (q^2) \nonumber \\
 X^{\ell'}_S & =  \frac{m^2_B -m^2_D}{m_b - m_c} C^{\ell'}_S \sqrt{q^2} F_0(q^2)\,,  \quad X^{\ell'}_T= - \frac{\sqrt{q^2 \lambda_D}}{m_B + m_D}  C^{\ell'}_T F_T(q^2) \nonumber \\
 C^{\ell'}_V & = \frac{\sqrt{2}}{8 G_F V_{cb} } \sum_a V_{2a}\left(\frac{\tilde{y}_{3\ell'} \tilde{y}_{a\ell' }}{ m^2_S} - \frac{y_{3\ell'} y_{a\ell' }}{2 m^2_\Delta} \right) \,, \nonumber \\
 C^{\ell'}_S & =  - \frac{\sqrt{2} }{4 G_F V_{cb} }  \left(\frac{\tilde{y}_{3\ell'} w_{2\ell'}}{2m^2_S}-\frac{k_{3\ell'} \tilde{k}_{\ell' 2 }}{2 m^2_\Phi }\right) \,, \quad 
 C^{\ell'}_T = \frac{\sqrt{2} }{16 G_F V_{cb} }  \left(\frac{\tilde{y}_{3\ell'} w_{2\ell'}}{2m^2_S}+\frac{k_{3\ell'} \tilde{k}_{\ell' 2 }}{2 m^2_\Phi }\right)\,,  \nonumber \\
 \lambda_H & = m^4_B + m^4_H + q^4 -2(m^2_B m^2_H + m^2_H q^2 + q^2 m^2_B) \,. \label{eq:Hfunc}
 \end{align}
 We note that the effective couplings $C_S^{\ell'}$ and $C_T^{\ell'}$ at the $m_b$ scale can be obtained from the LQ mass scale via the renormalization group (RG) equation. Our numerical analysis considers the RG running effects with $(C_S^{\ell'}/C_T^{\ell'})_{\mu=m_{b}}/(C_S^{\ell'}/C_T^{\ell'})_{\mu = \mathcal{O}({\rm TeV}) } \sim 2.0$ at the $m_b$ scale~\cite{Sakaki:2013bfa}. 
The $\bar B \to D^* \ell' \bar\nu_{\ell'}$ decays involve $D^*$ polarizations and more complicated transition form factors, so the differential decay rate determined by summing all of the $D^*$ helicities  can be written as:
 \begin{equation}
  \frac{d\Gamma^{\ell'}_{D^*}}{dq^2} =  \sum_{h=L,+,-} \frac{d\Gamma^{\ell' h}_{D^*}}{dq^2} = \frac{G^2_F |V_{cb}|^2   \sqrt{\lambda_{D^*}}}{256 \pi^3 m^3_B} \left( 1- \frac{m^2_{\ell'}}{q^2}\right)^2 \sum_{h=L,+,-} V^{\ell' h}_{D^*}(q^2)\,,  \label{eq:diffD*}
 \end{equation}
where $\lambda_{D^*}$ is found in Eq.~(\ref{eq:Hfunc}) and the detailed $\{ V^{\ell' h}_{D^*}\}$ functions are shown in the appendix. According to Eqs.~(\ref{eq:diffD}) and (\ref{eq:diffD*}),  $R_{M}$ ($M=D,D^*$) can be calculated by:
 \begin{equation}
 R_M = \frac{\int^{q^2_{\rm max}}_{m^2_\tau }dq^2 \left(d\Gamma^{\tau}_M/dq^2 \right)}{ \int^{q^2_{\rm max}}_{m^2_{\ell}} dq^2 \left( d\Gamma^\ell_M/dq^2 \right)}  \label{eq:R_M}
 \end{equation} 
where $q^2_{\rm max}=(m_B -m_M)^2$ and $\Gamma^{\ell}_M = (\Gamma^e_M + \Gamma^\mu_M)/2$.
 For the $B \to K \ell^+ \ell^-$ decays, the differential decay rate can be expressed as~\cite{Chen:2002zk}:
\begin{align}
\frac{d\Gamma_{K\ell \ell}(q^2)}{dq^2} & \approx  \frac{ |c_{\rm SM} |^2 m^3_B}{3\cdot 2^8 \pi^3} \left(1- \frac{q^2}{m^2_B}\right)^{3/2} \nonumber \\
& \times  \left[\left| C^\ell_9  F_+(q^2) + \frac{2 m_b C_7 }{m_B + m_K}  F_T(q^2)  \right|^2 + \left| C^\ell_{10}  F_+(q^2) \right|^2 \right]\,. \label{eq:diffBKll}
\end{align}
 From Eq.~(\ref{eq:diffBKll}), the measured ratio $R_K$ in the range $q^2=[q^2_{\rm min}, q^2_{\rm max}]=[1,6]$ GeV$^2$ can be estimated by:
\begin{equation}
R_K = \frac{\int^{q^2_{\rm max}}_{q^2_{\rm min}} dq^2 d\Gamma_{K\mu \mu}/dq^2 } {\int^{q^2_{\rm max}}_{q^2_{\rm min}}dq^2  
d\Gamma_{Kee}/dq^2 }\,. \label{eq:RK}
\end{equation}
 $R_{K^*}$ is similar to $R_K$, and thus we only show the result for $R_K$. 

\section{Numerical analysis}

After discussing the possible constraints and observables of interest, we now present the numerical analysis to determine the common parameter region where the $R_{D^{(*)}}$ and $R_{K^{(*)}}$ anomalies can fit the experimental data.
Before presenting the numerical analysis, we summarize the parameters involved, which are related to the specific measurements as follows:
 \begin{align}
 \text{muon}\ g-2  &:   k_{32} \tilde{k}_{23}\,, \tilde{y}_{32} w_{32}\,;\quad    R_{K} :  k_{3\ell} k_{2\ell}\,,~ y_{3\ell} y_{2\ell}  \,; \nonumber \\
 R_{D^{(*)}} &:  k_{3\ell'} \tilde{k}_{\ell' 2}\,,~ \sum_a V_{2a} \left( y_{3\ell' } y_{a\ell' }\,,~ \tilde{y}_{3\ell'} \tilde{y}_{a \ell'}\right)\,,~\tilde{y}_{3\ell'} w_{2\ell'}\,. \label{eq:pre}
  \end{align}
The parameters related to the radiative LFV, $\Delta B=2$, and $B^+\to K^+ \nu \bar\nu $ processes are defined as:
 \begin{align}
 \mu \to e \gamma &:  k_{32} \tilde{k}_{13}\,, \tilde{k}_{23} k_{31}\,, \tilde{y}_{32} w_{31}\,, w_{32} \tilde{y}_{31}\,; \nonumber \\
 \tau \to \ell_a \gamma &:  k_{33} \tilde{k}_{a 3}\,,\tilde{k}_{33} k_{3a}\,, \tilde{y}_{33} w_{3a}, w_{33}\tilde{y}_{3a}\,; \nonumber \\
 B^+ \to K^+ \nu \bar \nu &: \tilde{y}_{3i} \tilde{y}_{2i}\,, y_{3i} y_{2i} \,;   \quad B_s \to \mu^+ \mu^- :  k_{32} k_{22}\,,~y_{32}y_{22} \,;\nonumber \\
 \Delta m_{B_s} &: \left(\sum_i z_{3 i} z_{2i}\right)^2 \,,~
 \left( \sum_i  y_{3i} \tilde{y}_{2i} \right) \left(\sum_i \tilde{y}_{3i} y_{2i} \right) \,, \label{eq:con}
 \end{align}
 where $z_{3i}z_{2i} = k_{3i}k_{2i}, y_{3i} y_{2i}, \tilde{y}_{3i} \tilde{y}_{2i}$. From Eqs.~(\ref{eq:pre}) and (\ref{eq:con}), we can see that in order to avoid the $\mu \to e \gamma$ and $\tau \to \ell \gamma$ constraints and obtain a sizable and positive $\Delta a_\mu$, we can set ($\tilde{k}_{13,33}$, $k_{31,33}$, $w_{3i}$) as a small value. From the limit of $B^+ \to K^+ \nu \bar\nu$, we obtain $\tilde{y}_{3i} \tilde{y}_{2i}<0.03$, and thus the resulting $\Delta m_{B_s}$ is smaller than the current data. In order to further reduce the number of free parameters and avoid large fine-tuning couplings, we employ the scheme with $k_{ij} \approx \tilde{k}_{ji} \approx | y_{ij}|$, where the sign of $y_{ij}$ can be selected to obtain the correct sign for $C^{LQ,\ell_j}_{9}$ and to decrease the value of $C^{LQ, \mu}_{10}$ such that $B_s \to \mu^+ \mu^-$ can fit the experimental data. As mentioned earlier, to avoid the bounds from the $K$, $B_d$, and $D$ systems, we also use $k_{1\ell'} \approx \tilde{k}_{\ell' 1} \approx y_{1i} \approx \tilde{y}_{1i} \approx w_{1i} \sim 0$. When we omit these small couplings, the correlations of the parameters in Eqs.~(\ref{eq:pre}) and (\ref{eq:con}) can be simplified further as:
\begin{align}
 \text{muon}\ g-2  &:   k_{32}\tilde{k}_{23} \,; ~~R_{K} :  k_{32} k_{22}\,,~ ~y_{32} y_{22} \,;~ R_{D^{(*)}}  :  k_{32} k_{22}, ~y_{32} y_{22},~  \tilde{y}_{3 \ell'} w_{2\ell'}\,; \nonumber \\
 B_s \to \mu^+ \mu^- &: k_{32} k_{22}\,,~y_{32} y_{22}\,; ~~\Delta m_{B_s} :  (k_{3 2} k_{22})^2 \,, ~(y_{32} y_{22})^2\,,
  \label{eq:pre2}
  \end{align}
  where $\tilde{y}_{3i} \tilde{y}_{2i}$ are ignored due to the constraint from $B^+ \to K^+ \nu \bar\nu$.
The typical values of these parameters for fitting the anomalies in the $b\to s \mu^+ \mu^-$ decay are $y_{32}(k_{32}), y_{22}(k_{22}) \sim 0.07$, so the resulting $\Delta m_{B_s}$ is smaller than the current data, but these parameters are too small to explain $R_{D^{(*)}}$. Thus, we must depend on the singlet LQ to resolve the $R_{D}$ and $R_{D^*}$ excesses, where the main free parameters are now $\tilde{y}_{3\ell'} w_{2\ell'}$.

\begin{figure}[hptb] 
\includegraphics[width=70mm]{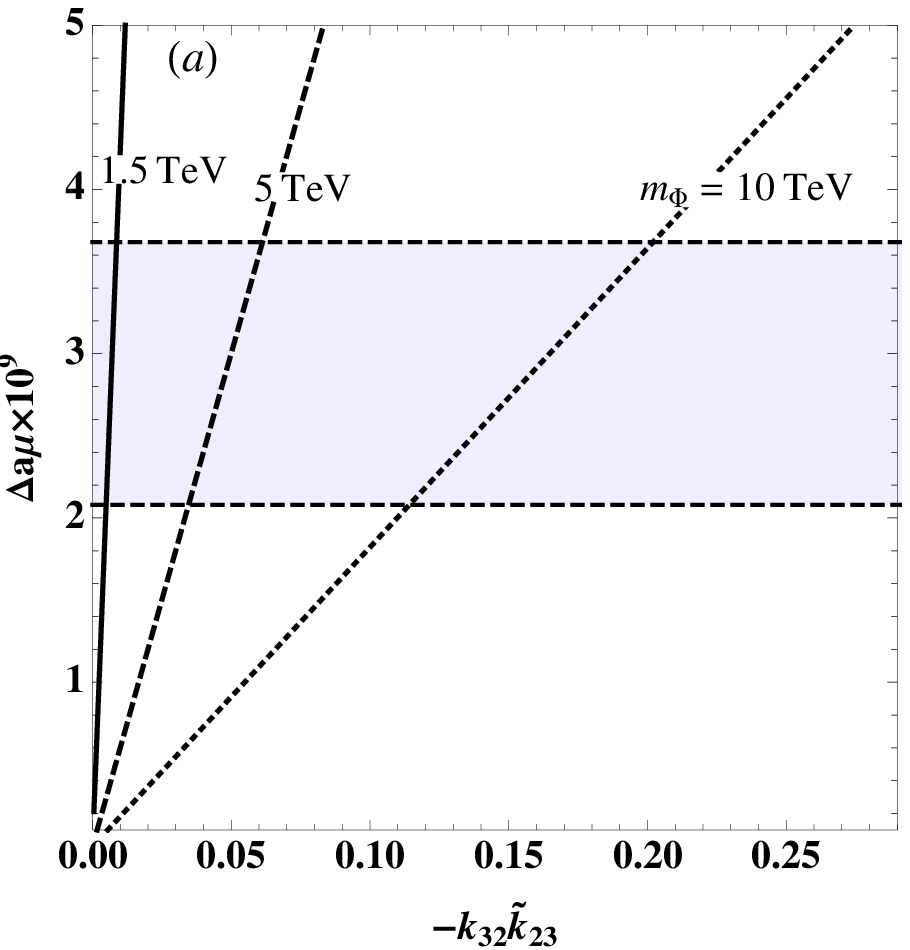} 
\includegraphics[width=70mm]{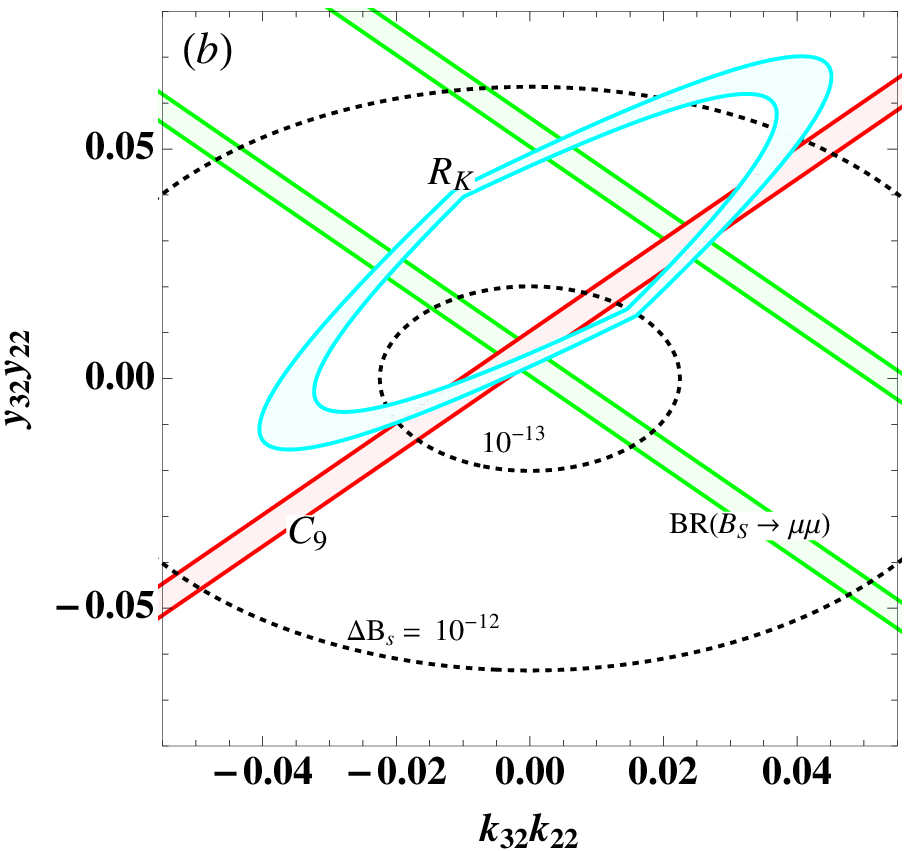} 
\caption{(a) $\Delta a_\mu$ as a function of $k_{32} \tilde{k}_{23}$ with $m_\Phi=1.5,\, 5,\, 10$ TeV, where the band denotes the experimental data with $1\sigma$ errors. (b) Contours for $R_K$, $B_s\to \mu^+ \mu^-$, $\Delta m_{B_s}$, and $C^{LQ,\mu}_9$ as a function of $k_{32}k_{22}$ and $y_{32}y_{22}$, where the ranges of $R_K$ and $B_s\to \mu^+ \mu^-$ are the experimental values with $1\sigma$ errors and $m_{LQ} = 1.5$ TeV. For $C^{LQ,\mu}_9$, we show the range for $C^{LQ,\mu}=[-1.5, -0.5]$.  }
\label{fig:gm2_RK}
\end{figure}

After discussing the constraints and the correlations among various processes, we present the numerical analysis in the following. There are several LQs in the model, but we use $m_{LQ}$ to denote the mass of a LQ. From Eqs.~(\ref{eq:aRL}), (\ref{eq:muong-2}), and (\ref{eq:pre2}), we can see that the muon $g-2$ depends only on $k_{32}\tilde{k}_{23}$ and $m_\Phi$. 
We illustrate $\Delta a_\mu$ as a function of $k_{32} \tilde{k}_{23}$ in Fig.~\ref{fig:gm2_RK}(a), where the solid, dashed, and dotted lines denote the results for $m_\Phi=1.5$, $5$, and $10$ TeV, respectively, and the band is the experimental value with $1\sigma$ errors. Due to the $m_t$ enhancement, $k_{32}\tilde{k}_{23} \sim 0.05$ with $m_\Phi \sim$ 1 TeV can explain the muon $g-2$ anomaly. 

According to the relationships shown in Eq.~(\ref{eq:pre2}), $R_K$, $B_s\to \mu^+ \mu^-$, and $\Delta m_{B_s}$ depend on the same parameters, i.e., $k_{32}k_{22}$ and $y_{32} y_{22}$. We show the contours for these observables as a function of $k_{32}k_{22}$ and $y_{32}y_{22}$ in Fig.~\ref{fig:gm2_RK}(b), where the data with $1\sigma$ errors and $m_{LQ}=1.5$ TeV are taken for all LQ masses. Based on these results, we see clearly that $\Delta m_{B_s} < \Delta m^{\rm exp}_{B_s}$ in the range of $|k_{32}k_{22}|$, $|y_{32}y_{22}| < 0.05$, where $R_K$ and $BR(B_s\to \mu^+ \mu^-)$ can both fit the experimental data simultaneously. In addition, we show $C^{LQ,\mu}_{9}=[-1.5,-0.5]$ in the same plot. We can see that $C^{LQ,\mu}_9 \sim -1$, which is used to explain the angular observable $P'_5$, can also be achieved in the same common region. According to Fig.~\ref{fig:gm2_RK}(b), the preferred values of $k_{32}k_{22}$ and $y_{32}y_{22}$ where the observed $R_K$ and $B_s\to \mu^+ \mu^-$ and the $C^{LQ,\mu}_9=[-1.5, -0.5]$ overlap are around $( k_{32}k_{22}, y_{32}y_{22} ) \sim ( -0.001,\, 0.004 )$ and $\sim ( 0.025,\, 0.03 )$. The latter values are at the percentage level but they are still not sufficiently large to explain the tree-dominated $R_D$ and $R_{D^*}$ anomalies.

\begin{figure}[hptb] 
\includegraphics[width=70mm]{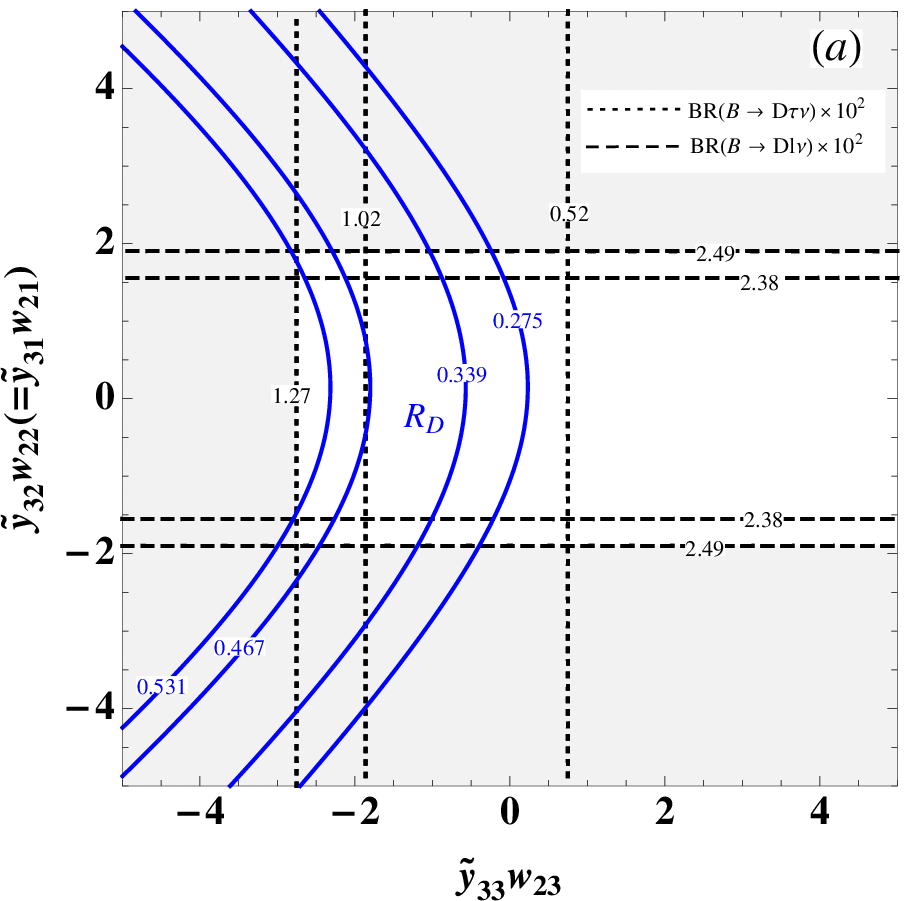} 
\includegraphics[width=70mm]{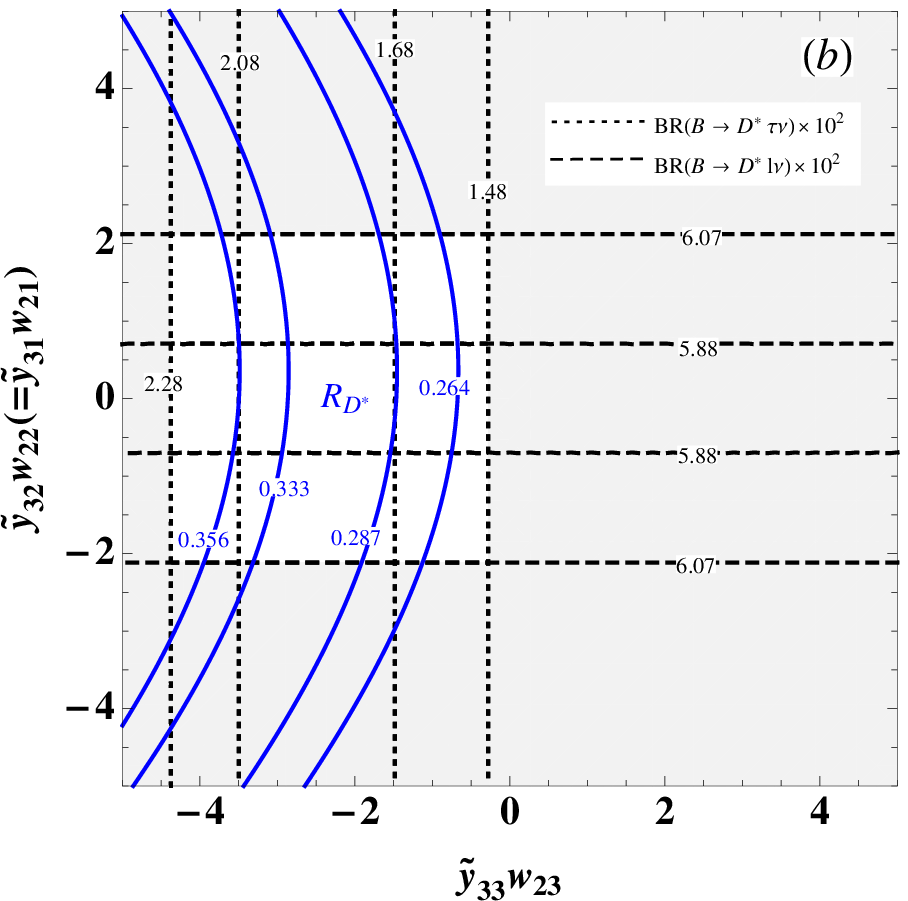} 
\includegraphics[width=70mm]{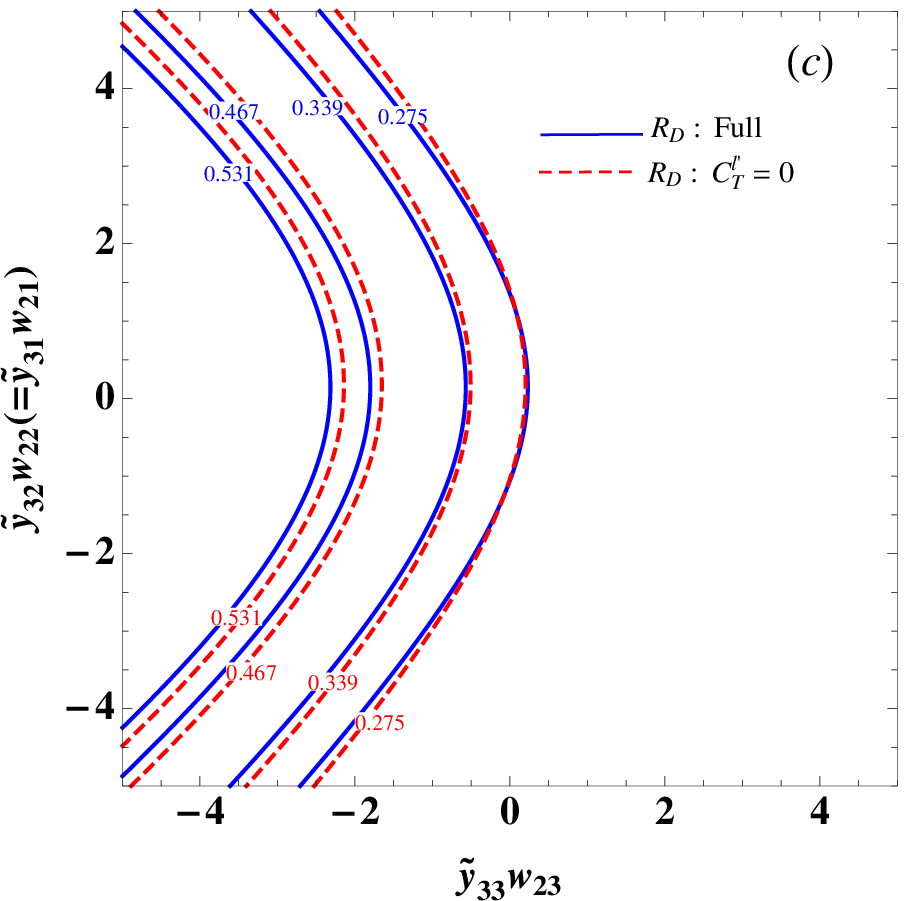} 
\includegraphics[width=70mm]{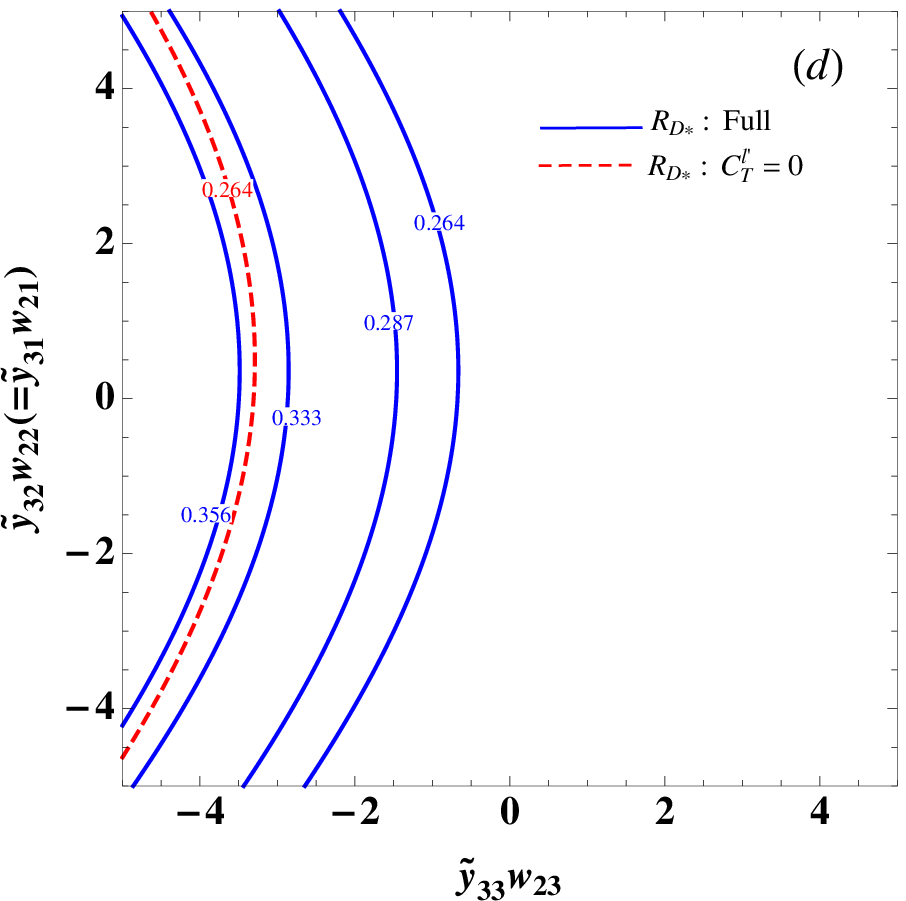} 
\caption{ Contours for (a) $R_{D}$ and (b) $R_{D^*}$, where the solid lines denote the data with $1\sigma$ and $2\sigma$ errors, respectively. The horizontal dashed lines in both plots denote the $BR^{\rm exp}(B^+ \to D^{(*)} \ell \nu_\ell)$ whereas the vertical dotted lines are the $BR^{\rm exp}(B^+\to D^{(*)} \tau \nu_\tau)$. 
 Contours for (c) $R_{D}$ and (d) $R_{D^*}$, where the solid and dashed lines denote the situations with and without tensor operator contributions, respectively. In this case, we take $m_{LQ} = 1.5$ TeV. }
\label{fig:RD_RDv}
\end{figure}
\begin{figure}[tbph] 
\includegraphics[width=85mm]{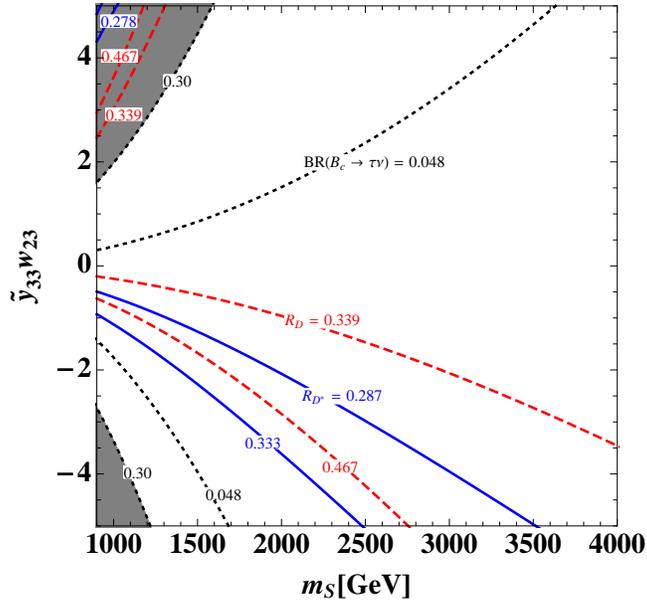} 
\caption{ Contours for $BR(B_c \to \tau \bar \nu_\tau)$ and $R_{D^{(*)}}$ as a function of $w_{23} \tilde y_{23}$ and $m_S$. }
\label{fig:Bc}
\end{figure}

After studying the muon $g-2$ and $R_K$ anomalies, we numerically analyze the ratio of $BR(\bar B\to D^{(*)} \tau \bar\nu_\tau)$ to $BR(\bar B \to D^{(*)} \ell \bar \nu_\ell)$, i.e., $R_{D^(*)}$. The introduced doublet and triplet LQs cannot efficiently enhance $R_{D^{(*)}}$, so in the following estimations, we only focus on the singlet LQ contributions, where the four-Fermi interactions shown in Eq.~(\ref{eq:Hbc}) come mainly from the scalar- and tensor-type interaction structures. Based on Eqs.~(\ref{eq:diffD}), (\ref{eq:diffD*}), and (\ref{eq:R_M}), we show the contours for $R_{D}$ and $R_{D^*}$ as a function of $\tilde{y}_{33} w_{23}$ and $\tilde{y}_{32} w_{22}(\tilde{y}_{31} w_{21})$ in Fig.~\ref{fig:RD_RDv}(a) and (b), where the horizontal dashed and vertical dotted lines in both plots denote $BR^{\rm exp}(B^- \to D [\ell \bar \nu_\ell,  \tau \bar\nu_\tau])=[2.27 \pm 0.11,\, 0.77 \pm 0.25]\%$ and $BR^{\rm exp}(B^- \to D^{*} [\ell \nu_\ell,\, \tau \bar\nu_\tau])=[5.69\pm 0.19, \, 1.88\pm 0.20]\%$, respectively, and $m_{LQ} = 1.5$ TeV is used, and the data with $2\sigma$ errors are taken. For simplicity, we take $\tilde{y}_{31} w_{21}\approx \tilde{y}_{32} w_{22}$. When considering the limits from $BR(\bar B\to D^{(*)} \ell' \bar\nu_{\ell'})$, we obtain the limits $|\tilde{y}_{3\ell}w_{2\ell}| \leq 1.5$ and $\tilde{y}_{33} w_{23} > 0$. In order to clearly demonstrate the influence of tensor-type interactions, we also calculate the situation by setting $C^{\ell'}_T=0$. The contours obtained for $R_D$ and $R_{D^*}$ are shown in Fig.~\ref{fig:RD_RDv}(c) and (d), where the solid and dashed lines denote the cases with and without $C^{\ell'}_{T}$, respectively. According to these plots, we can see that $R_D$ and $R_{D^*}$ have different responses to the tensor operators, where the latter is more sensitive to the tensor interactions. $R_D$ and $R_{D^*}$ can be explained simultaneously with the tensor couplings.
In order to understand the correlation between $BR(B_c \to \tau \bar \nu_\tau)$ and $R_{D^{(*)}}$, we show the contours for $BR(B_c \to \tau \bar \nu_\tau)$ and $R_{D^{(*)}}$ as a function of $w_{23} \tilde y_{33}$ and $m_{S}$ in Fig.~\ref{fig:Bc}, where $\tilde y_{32} w_{22} \approx \tilde y_{31} w_{21}\approx 0$ are used, and the gray area is excluded by $BR(B_c^- \to \tau \nu ) < 0.3$. We can see that the predicted $BR(B_c \to \tau \bar \nu_\tau)$ is much smaller than the experimental bound.  

Finally, we make some remarks regarding the constraint due to the LQ search at the LHC. Due to the flavor physics constraints, only the $S^{1/3}$ Yukawa couplings $\tilde y_{t \tau }$, $\tilde y_{b\nu_{\tau}}$, and $w_{c\tau}$ can be of ${\cal O}(1)$. These couplings affect the $S^{1/3}$ decays but also their production. Therefore, in addition to the $S^{1/3}$-pair production, based on the $O(1)$ Yukawa couplings, the single $S^{1/3}$ production becomes interesting. In the $pp$ collisions, the single $S^{1/3}$ production can be generated via the $g b \to S^{-1/3} \bar \nu_\tau$ and $gc \to S^{-1/3} \tau^+$ channels. Using CalcHEP 3.6~\cite{Belyaev:2012qa,Belyaev:2005ew} with the CTEQ6 parton distribution functions~\cite{Nadolsky:2008zw}, their production cross sections with $|w_{23}|\sim |\tilde y_{b\nu_\tau}| \sim \sqrt{2}$ and $m_{LQ}=1000$ GeV at $\sqrt{s}=13$ TeV can be obtained as 3.9 fb and 2.9 fb, respectively, whereas the $S^{1/3}$-pair production cross section is $\sigma(pp\to S^{-1/3} S^{1/3})\approx 2.4$ fb. If we assume that $S^{-1/3}$ predominantly decays into $ t \tau$, $b \nu_\tau$, and $ c \tau$ with similar BRs, i.e. $BR(S^{-1/3}\to f)\sim 1/3$, then the single $S^{1/3}$ production cross section $\sigma( S^{-1/3} X)$ times $BR(S^{-1/3}\to f)$ with $X$ and $f$ as the possible final states can be estimated as around $1$ fb. The LQ coupling $w_{23}$ involves different generations, so the constraints due to the collider measurements may not be applied directly. However, if we compare this with the CMS experiment~\cite{Khachatryan:2015qda} based on a single production of the second-generation scalar LQ, we find that the values of $\sigma \times BR$ at $m_{LQ} \sim 1000$ GeV are still lower than the CMS upper limit with few fb. The significance of this discovery depends on the kinematic cuts and event selection conditions, but this discussion is beyond the scope of this study and we leave the detailed analysis for future research.

\section{Summary}

In this study, we considered the muon $g-2$, $R_{K^{(*)}}$, and $R_{D^{(*)}}$ anomalies in a specific model with one doublet, triplet, and singlet LQ. We demonstrated that the muon $g-2$ can be explained only by the doublet LQ due to the $m_t$ enhancement. The combined effects of the doublet and triplet LQs can lead to $C^{LQ,\mu}_9 \sim -1$, which can resolve the $R_{K^{(*)}}$ anomaly and the excess of the angular observable $P'_5$ in the $B\to K^* \mu^+ \mu^-$ decays. When we considered the constraints due to $\ell' \to \ell \gamma$, $\Delta m_{B_s}$, $BR(B^+\to K^+ \nu \bar\nu)$, $BR(\bar B\to D^{(*)} \ell' \bar\nu_{\ell'})$, and $BR(B_c \to \tau \bar \nu_\tau)$, we found that the singlet LQ contributions can enhance $R_{D}$ and $R_{D^*}$ would be consistent with the current measurements obtained through the scalar and tensor four-Fermi interactions. We also found that $R_{D^*}$ is not sensitive to scalar interactions but it is sensitive to the tensor interactions, although the influence on $R_{D}$ is reversed. Using the LQ Yukawa couplings of $O(1)$, we estimated the single production cross section of the scalar LQ and its decaying BRs, where the results are still under the CMS upper limit. The significance of this discovery requires validation in a detailed event simulation, which is beyond the scope of the present study. 

\section*{Acknowledgments}

This study was partially supported by the Ministry of Science and Technology of Taiwan  under grant MOST-103-2112-M-006-004-MY3 (CHC).

\appendix
\section*{ Appendix}

The helicity-dependent functions, $V^{\ell' L, + ,-}_{D^*}$ are defined as follows. For longitudinal polarization, the $V^{\ell' L}_{D^*}$ function is defined as:
\begin{align}
V^{\ell' L}_{D^*}(q^2) & = \frac{2}{3} \left(2+ \frac{m^2_{\ell'}}{q^2} \right) |h_0|^2 + \frac{2}{3} \left(1+ 2\frac{m^2_{\ell'}}{q^2} \right)|h^0_T|^2 \nonumber \\
& + \frac{2 m^2_{\ell'}}{q^2} \lambda_{D^*} A^2_0 \left|1+C^{\ell'}_V + \frac{q^2\, C^{\ell'}_S}{m_{\ell'} (m_b + m_c)}  \right|^2 - \frac{16 m_{\ell'}}{\sqrt{q}} h_{0} h^0_{T}
\end{align}
with
\begin{align}
h_0 &= \frac{1+C^{\ell'}_V}{2m_{D^*}} \left( (m^2_B -m^2_{D^*} -q^2) (m_B + m_{D^*}) A_1 - \frac{\lambda_{D^*}}{m_B + m_{D^*}} A_2\right)  \,, \nonumber \\
h^0_T&= \frac{C^{\ell'}_T \sqrt{q^2} }{2m_{D^*}} \left( (m^2_B + 3 m^2_{D^*} -q^2) T_2 - \frac{\lambda_{D^*}}{m^2_B - m^2_{D^*}} T_3\right) \,,
\end{align}
where $C^{\ell'}_{V,S,T}$ denote the LQ contributions, which can be found in Eq.~(\ref{eq:Hfunc}), and  the dependence of $q^2$ on the form factors is suppressed. For transverse polarizations, the $V^{\ell' \pm}_{D^*}$ functions are defined as:
\begin{align}
V^{\ell' \pm}_{D^*}(q^2) &= \frac{2 q^2 }{3} \left( 2+ \frac{m^2_{\ell'}}{q^2}\right) |h_{\pm }|^2 + \frac{32 q^2 }{3} \left( 1 + \frac{2 m^2_{\ell'}}{q^2}\right) |h^{\pm}_{T}|^2 - 16 m_{\ell'} \sqrt{q^2} h_{\pm} h^{\pm}_{ T}  \,, \\
h_{\pm} &=(1+ C^{\ell'}_V) \left[(m_B+m_{D^*})A_1 \mp \frac{\sqrt{\lambda_{D^*}} }{m_B+m_{D^*}}V \right]\,, \nonumber \\
h^{\pm}_{ T} & = \frac{C^{\ell'}_T}{\sqrt{q^2}} \left[  \left( m^2_B - m^2_{D^*} \right)T_2  \pm \sqrt{\lambda_{D^*}} T_1 \right] \,.\nonumber 
\end{align}


\begin{thebibliography}{99}


   \bibitem{DescotesGenon:2012zf} 
  S.~Descotes-Genon, J.~Matias, M.~Ramon, and J.~Virto,
  JHEP {\bf 1301}, 048 (2013)
  [arXiv:1207.2753 [hep-ph]].


\bibitem{Aaij:2015oid} 
  R.~Aaij {\it et al.} [LHCb Collaboration],
  JHEP {\bf 1602}, 104 (2016)
  [arXiv:1512.04442 [hep-ex]].


\bibitem{Aaij:2013qta} 
  R.~Aaij {\it et al.} [LHCb Collaboration],
  Phys.\ Rev.\ Lett.\  {\bf 111}, 191801 (2013)
  [arXiv:1308.1707 [hep-ex]].
  

  
\bibitem{Wehle:2016yoi} 
  S.~Wehle {\it et al.} [Belle Collaboration],
  arXiv:1612.05014 [hep-ex].
  
\bibitem{Huschle:2015rga} 
  M.~Huschle {\it et al.} [Belle Collaboration],
  Phys.\ Rev.\ D {\bf 92}, no. 7, 072014 (2015)
  [arXiv:1507.03233 [hep-ex]].
  
\bibitem{Lees:2012xj} 
  J.~P.~Lees {\it et al.} [BaBar Collaboration],
  Phys.\ Rev.\ Lett.\  {\bf 109}, 101802 (2012)
  [arXiv:1205.5442 [hep-ex]].
  
\bibitem{Lees:2013uzd} 
  J.~P.~Lees {\it et al.} [BaBar Collaboration],
  Phys.\ Rev.\ D {\bf 88}, no. 7, 072012 (2013)
  [arXiv:1303.0571 [hep-ex]].
  
\bibitem{Abdesselam:2016cgx} 
  A.~Abdesselam {\it et al.} [Belle Collaboration],
  arXiv:1603.06711 [hep-ex].
  
\bibitem{Hirose:2016wfn} 
  S.~Hirose {\it et al.} [Belle Collaboration],
  arXiv:1612.00529 [hep-ex].

  
\bibitem{Aaij:2015yra} 
  R.~Aaij {\it et al.} [LHCb Collaboration],
  Phys.\ Rev.\ Lett.\  {\bf 115}, no. 11, 111803 (2015)
  Addendum: [Phys.\ Rev.\ Lett.\  {\bf 115}, no. 15, 159901 (2015)]
  [arXiv:1506.08614 [hep-ex]].

\bibitem{Amhis:2016xyh} 
  Y.~Amhis {\it et al.},
  arXiv:1612.07233 [hep-ex].


\bibitem{Lattice:2015rga}
  J.~A.~Bailey {\it et al.} [MILC Collaboration],
  Phys.\ Rev.\ D {\bf 92} (2015) no.3,  034506
  [arXiv:1503.07237 [hep-lat]].
 
 \bibitem{Na:2015kha} 
  H.~Na {\it et al.} [HPQCD Collaboration],
  Phys.\ Rev.\ D {\bf 92}, no. 5, 054510 (2015)
  Erratum: [Phys.\ Rev.\ D {\bf 93}, no. 11, 119906 (2016)]
  [arXiv:1505.03925 [hep-lat]].
  

  

\bibitem{Aaij:2014ora} 
  R.~Aaij {\it et al.} [LHCb Collaboration],
  Phys.\ Rev.\ Lett.\  {\bf 113}, 151601 (2014)
  [arXiv:1406.6482 [hep-ex]].
  
\bibitem{Aaij:2017vbb} 
  R.~Aaij {\it et al.} [LHCb Collaboration],
  arXiv:1705.05802 [hep-ex].


      \bibitem{PDG} C. Patrignani et al. (Particle Data Group), Chin. Phys. C, {\bf 40}, 100001 (2016).   

  
  
    \bibitem{Matias:2012xw} 
  J.~Matias, F.~Mescia, M.~Ramon, and J.~Virto,
  JHEP {\bf 1204}, 104 (2012)
  [arXiv:1202.4266 [hep-ph]].
  
    \bibitem{Fajfer:2012vx} 
  S.~Fajfer, J.~F.~Kamenik, and I.~Nisandzic,
  Phys.\ Rev.\ D {\bf 85}, 094025 (2012)
  [arXiv:1203.2654 [hep-ph]].
  
\bibitem{Fajfer:2012jt} 
  S.~Fajfer, J.~F.~Kamenik, I.~Nisandzic, and J.~Zupan,
  Phys.\ Rev.\ Lett.\  {\bf 109}, 161801 (2012)
  [arXiv:1206.1872 [hep-ph]].
  
\bibitem{Crivellin:2012ye} 
  A.~Crivellin, C.~Greub, and A.~Kokulu,
  Phys.\ Rev.\ D {\bf 86}, 054014 (2012)
  [arXiv:1206.2634 [hep-ph]].

\bibitem{Datta:2012qk} 
  A.~Datta, M.~Duraisamy, and D.~Ghosh,
  Phys.\ Rev.\ D {\bf 86}, 034027 (2012)
  [arXiv:1206.3760 [hep-ph]].

\bibitem{Deshpande:2012rr} 
  N.~G.~Deshpande and A.~Menon,
  JHEP {\bf 1301}, 025 (2013)
  [arXiv:1208.4134 [hep-ph]].
  
\bibitem{He:2012zp} 
  X.~G.~He and G.~Valencia,
  Phys.\ Rev.\ D {\bf 87}, no. 1, 014014 (2013)
  [arXiv:1211.0348 [hep-ph]].
  
\bibitem{Tanaka:2012nw} 
  M.~Tanaka and R.~Watanabe,
  Phys.\ Rev.\ D {\bf 87}, no. 3, 034028 (2013)
  [arXiv:1212.1878 [hep-ph]].
  
\bibitem{Ko:2012sv} 
  P.~Ko, Y.~Omura and C.~Yu,
  JHEP {\bf 1303}, 151 (2013)
  [arXiv:1212.4607 [hep-ph]].

\bibitem{Dorsner:2013tla} 
  I.~Dorsner, S.~Fajfer, N.~Kosnik, and I.~Nisandzic,
  JHEP {\bf 1311}, 084 (2013)
  [arXiv:1306.6493 [hep-ph]].


\bibitem{Descotes-Genon:2013wba} 
  S.~Descotes-Genon, J.~Matias, and J.~Virto,
  Phys.\ Rev.\ D {\bf 88}, 074002 (2013)
  [arXiv:1307.5683 [hep-ph]].

\bibitem{Sakaki:2013bfa} 
  Y.~Sakaki, M.~Tanaka, A.~Tayduganov, and R.~Watanabe,
  Phys.\ Rev.\ D {\bf 88}, no. 9, 094012 (2013)
  [arXiv:1309.0301 [hep-ph]].
  
\bibitem{Gauld:2013qja} 
  R.~Gauld, F.~Goertz, and U.~Haisch,
  JHEP {\bf 1401}, 069 (2014)
  [arXiv:1310.1082 [hep-ph]].

\bibitem{Datta:2013kja} 
  A.~Datta, M.~Duraisamy, and D.~Ghosh,
  Phys.\ Rev.\ D {\bf 89}, no. 7, 071501 (2014)
  [arXiv:1310.1937 [hep-ph]].
  
\bibitem{Abada:2013aba} 
  A.~Abada, A.~M.~Teixeira, A.~Vicente, and C.~Weiland,
  JHEP {\bf 1402}, 091 (2014)
  [arXiv:1311.2830 [hep-ph]].


\bibitem{Hurth:2013ssa} 
  T.~Hurth and F.~Mahmoudi,
  JHEP {\bf 1404}, 097 (2014)
  [arXiv:1312.5267 [hep-ph]].
  
\bibitem{Duraisamy:2014sna} 
  M.~Duraisamy, P.~Sharma, and A.~Datta,
  Phys.\ Rev.\ D {\bf 90}, no. 7, 074013 (2014)
  [arXiv:1405.3719 [hep-ph]].

\bibitem{Descotes-Genon:2014uoa} 
  S.~Descotes-Genon, L.~Hofer, J.~Matias, and J.~Virto,
  JHEP {\bf 1412}, 125 (2014)
  [arXiv:1407.8526 [hep-ph]].
 
\bibitem{Altmannshofer:2014rta} 
  W.~Altmannshofer and D.~M.~Straub,
  Eur.\ Phys.\ J.\ C {\bf 75}, no. 8, 382 (2015)
  [arXiv:1411.3161 [hep-ph]].
  
\bibitem{Crivellin:2015mga} 
  A.~Crivellin, G.~D'Ambrosio, and J.~Heeck,
  Phys.\ Rev.\ Lett.\  {\bf 114}, 151801 (2015)
  [arXiv:1501.00993 [hep-ph]].
  
  

\bibitem{Sahoo:2015wya} 
  S.~Sahoo and R.~Mohanta,
  Phys.\ Rev.\ D {\bf 91}, no. 9, 094019 (2015)
  [arXiv:1501.05193 [hep-ph]].
  
\bibitem{Straub:2015ica} 
  A.~Bharucha, D.~M.~Straub, and R.~Zwicky,
  arXiv:1503.05534 [hep-ph].
  
\bibitem{Becirevic:2015asa} 
  D.~Becirevic, S.~Fajfer, and N.~Kosnik,
  Phys.\ Rev.\ D {\bf 92}, no. 1, 014016 (2015)
  [arXiv:1503.09024 [hep-ph]].

\bibitem{Crivellin:2015era} 
  A.~Crivellin, L.~Hofer, J.~Matias, U.~Nierste, S.~Pokorski, and J.~Rosiek,
  Phys.\ Rev.\ D {\bf 92}, no. 5, 054013 (2015)
  [arXiv:1504.07928 [hep-ph]].
  
\bibitem{Lee:2015qra} 
  C.~J.~Lee and J.~Tandean,
  JHEP {\bf 1508}, 123 (2015)
  [arXiv:1505.04692 [hep-ph]].
  
\bibitem{Alonso:2015sja} 
  R.~Alonso, B.~Grinstein, and J.~Martin Camalich,
  JHEP {\bf 1510}, 184 (2015)
  [arXiv:1505.05164 [hep-ph]].
  
\bibitem{Freytsis:2015qca} 
  M.~Freytsis, Z.~Ligeti, and J.~T.~Ruderman,
  Phys.\ Rev.\ D {\bf 92}, no. 5, 054018 (2015)
  [arXiv:1506.08896 [hep-ph]].
  
\bibitem{Sahoo:2015qha} 
  S.~Sahoo and R.~Mohanta,
  Phys.\ Rev.\ D {\bf 93}, no. 3, 034018 (2016)
  [arXiv:1507.02070 [hep-ph]].
  

\bibitem{Crivellin:2015hha} 
  A.~Crivellin, J.~Heeck, and P.~Stoffer,
  Phys.\ Rev.\ Lett.\  {\bf 116}, no. 8, 081801 (2016)
  [arXiv:1507.07567 [hep-ph]].

\bibitem{Descotes-Genon:2015uva} 
  S.~Descotes-Genon, L.~Hofer, J.~Matias, and J.~Virto,
  JHEP {\bf 1606}, 092 (2016)
  [arXiv:1510.04239 [hep-ph]].
   
\bibitem{Fajfer:2015ycq} 
  S.~Fajfer and N.~Kosnik,
  Phys.\ Lett.\ B {\bf 755}, 270 (2016)
  [arXiv:1511.06024 [hep-ph]].
   
\bibitem{Sahoo:2015pzk} 
  S.~Sahoo and R.~Mohanta,
  Phys.\ Rev.\ D {\bf 93}, no. 11, 114001 (2016)
  [arXiv:1512.04657 [hep-ph]].
  
\bibitem{Chiang:2016qov} 
  C.~W.~Chiang, X.~G.~He, and G.~Valencia,
  Phys.\ Rev.\ D {\bf 93}, no. 7, 074003 (2016)
  [arXiv:1601.07328 [hep-ph]].
  
  
\bibitem{Dorsner:2016wpm} 
  I.~Dorsner, S.~Fajfer, A.~Greljo, J.~F.~Kamenik, and N.~Kosnik,
  Phys.\ Rept.\  {\bf 641}, 1 (2016)
  [arXiv:1603.04993 [hep-ph]].
  
  \bibitem{Boucenna:2016wpr} 
  S.~M.~Boucenna, A.~Celis, J.~Fuentes-Martin, A.~Vicente, and J.~Virto,
  Phys.\ Lett.\ B {\bf 760}, 214 (2016)
  [arXiv:1604.03088 [hep-ph]].

\bibitem{Ivanov:2016qtw} 
  M.~A.~Ivanov, J.~G.~Korner, and C.~T.~Tran,
  Phys.\ Rev.\ D {\bf 94}, no. 9, 094028 (2016)
  [arXiv:1607.02932 [hep-ph]].


\bibitem{Hiller:2016kry} 
  G.~Hiller, D.~Loose, and K.~Schonwald,
  arXiv:1609.08895 [hep-ph].
  
\bibitem{Cheung:2016fjo} 
  K.~Cheung, T.~Nomura, and H.~Okada,
  Phys.\ Rev.\ D {\bf 94}, no. 11, 115024 (2016)
  [arXiv:1610.02322 [hep-ph]].
  
  \bibitem{Bardhan:2016uhr} 
  D.~Bardhan, P.~Byakti, and D.~Ghosh,
  JHEP {\bf 1701}, 125 (2017)
  [arXiv:1610.03038 [hep-ph]].

  
\bibitem{Cheung:2016frv} 
  K.~Cheung, T.~Nomura, and H.~Okada,
  Phys.\ Rev.\ D {\bf 95}, no. 1, 015026 (2017)
  [arXiv:1610.04986 [hep-ph]].
  
\bibitem{Wang:2016ggf} 
  L.~Wang, J.~M.~Yang, and Y.~Zhang,
  arXiv:1610.05681 [hep-ph].
  
\bibitem{ColuccioLeskow:2016dox} 
  E.~Coluccio Leskow, A.~Crivellin, G.~D'Ambrosio, and D.~Muller,
  arXiv:1612.06858 [hep-ph].

  
\bibitem{Cheung:2017efc} 
  K.~Cheung, T.~Nomura, and H.~Okada,
  arXiv:1701.01080 [hep-ph].
  
\bibitem{Ivanov:2017mrj} 
  M.~A.~Ivanov, J.~G.~Korner, and C.~T.~Tran,
  Phys.\ Rev.\ D {\bf 95}, no. 3, 036021 (2017)
  [arXiv:1701.02937 [hep-ph]].
  
\bibitem{Ko:2017quv} 
  P.~Ko, T.~Nomura, and H.~Okada,
  arXiv:1701.05788 [hep-ph].
  
  
\bibitem{Wei:2017ago} 
  M.~Wei and Y.~Chong-Xing,
  Phys.\ Rev.\ D {\bf 95}, no. 3, 035040 (2017)
  [arXiv:1702.01255 [hep-ph]].
  
  
\bibitem{Ko:2017yrd} 
  P.~Ko, T.~Nomura, and H.~Okada,
  arXiv:1702.02699 [hep-ph].
  
  
\bibitem{Cvetic:2017gkt} 
  G.~Cvetic, F.~Halzen, C.~S.~Kim, and S.~Oh,
  arXiv:1702.04335 [hep-ph].


    \bibitem{Hiller:2014yaa} 
  G.~Hiller and M.~Schmaltz,
  Phys.\ Rev.\ D {\bf 90}, 054014 (2014)
  [arXiv:1408.1627 [hep-ph]].


\bibitem{Hurth:2014vma} 
  T.~Hurth, F.~Mahmoudi, and S.~Neshatpour,
  JHEP {\bf 1412}, 053 (2014)
  [arXiv:1410.4545 [hep-ph]].
  

\bibitem{Glashow:2014iga} 
  S.~L.~Glashow, D.~Guadagnoli, and K.~Lane,
  Phys.\ Rev.\ Lett.\  {\bf 114}, 091801 (2015)
  [arXiv:1411.0565 [hep-ph]].
  
\bibitem{Gripaios:2014tna} 
  B.~Gripaios, M.~Nardecchia, and S.~A.~Renner,
  JHEP {\bf 1505}, 006 (2015)
  [arXiv:1412.1791 [hep-ph]].

  \bibitem{Sahoo:2015fla} 
  S.~Sahoo and R.~Mohanta,
  New J.\ Phys.\  {\bf 18}, no. 1, 013032 (2016)
  [arXiv:1509.06248 [hep-ph]].


\bibitem{Bauer:2015knc} 
  M.~Bauer and M.~Neubert,
  Phys.\ Rev.\ Lett.\  {\bf 116}, no. 14, 141802 (2016)
  [arXiv:1511.01900 [hep-ph]].
  
  
\bibitem{Das:2016vkr} 
  D.~Das, C.~Hati, G.~Kumar, and N.~Mahajan,
  Phys.\ Rev.\ D {\bf 94}, no. 5, 055034 (2016)
  [arXiv:1605.06313 [hep-ph]].

\bibitem{Li:2016vvp} 
  X.~Q.~Li, Y.~D.~Yang, and X.~Zhang,
  JHEP {\bf 1608}, 054 (2016)
  [arXiv:1605.09308 [hep-ph]].

\bibitem{Chen:2016dip} 
  C.~H.~Chen, T.~Nomura, and H.~Okada,
  Phys.\ Rev.\ D {\bf 94}, no. 11, 115005 (2016)
  [arXiv:1607.04857 [hep-ph]].

\bibitem{Becirevic:2016oho} 
  D.~Becirevic, N.~Kosnik, O.~Sumensari, and R.~Zukanovich Funchal,
  JHEP {\bf 1611}, 035 (2016)
  [arXiv:1608.07583 [hep-ph]].
  
  
\bibitem{Becirevic:2016yqi} 
  D.~Becirevic, S.~Fajfer, N.~Kosnik, and O.~Sumensari,
  Phys.\ Rev.\ D {\bf 94}, no. 11, 115021 (2016)
  [arXiv:1608.08501 [hep-ph]].
  
\bibitem{Sahoo:2016pet} 
  S.~Sahoo, R.~Mohanta, and A.~K.~Giri,
  arXiv:1609.04367 [hep-ph].


\bibitem{Bhattacharya:2016mcc} 
  B.~Bhattacharya, A.~Datta, J.~P.~Guevin, D.~London, and R.~Watanabe,
  arXiv:1609.09078 [hep-ph].

\bibitem{Duraisamy:2016gsd} 
  M.~Duraisamy, S.~Sahoo, and R.~Mohanta,
  arXiv:1610.00902 [hep-ph].
  


  
\bibitem{Dowdall:2013tga} 
  R.~J.~Dowdall {\it et al.} [HPQCD Collaboration],
  Phys.\ Rev.\ Lett.\  {\bf 110}, no. 22, 222003 (2013)
  [arXiv:1302.2644 [hep-lat]].
  


  
\bibitem{Buchalla:1995vs} 
  G.~Buchalla, A.~J.~Buras, and M.~E.~Lautenbacher,
  Rev.\ Mod.\ Phys.\  {\bf 68}, 1125 (1996)
  [hep-ph/9512380].
  
  
\bibitem{Aaij:2017vad} 
  R.~Aaij {\it et al.} [LHCb Collaboration],
  Phys.\ Rev.\ Lett.\  {\bf 118}, no. 19, 191801 (2017)
  doi:10.1103/PhysRevLett.118.191801
  [arXiv:1703.05747 [hep-ex]].

\bibitem{Bobeth:2013uxa} 
  C.~Bobeth, M.~Gorbahn, T.~Hermann, M.~Misiak, E.~Stamou, and M.~Steinhauser,
  Phys.\ Rev.\ Lett.\  {\bf 112}, 101801 (2014)
  [arXiv:1311.0903 [hep-ph]].

\bibitem{Melikhov:2000yu} 
  D.~Melikhov and B.~Stech,
  Phys.\ Rev.\ D {\bf 62}, 014006 (2000)
  [hep-ph/0001113].
  
\bibitem{Wang:2017jow} 
  Y.~M.~Wang, Y.~B.~Wei, Y.~L.~Shen, and C.~D.~Lu,
  arXiv:1701.06810 [hep-ph].

\bibitem{Chen:2002zk} 
  C.~H.~Chen and C.~Q.~Geng,
  Phys.\ Rev.\ D {\bf 66}, 094018 (2002)
  [hep-ph/0209352].
  
\bibitem{Alonso:2016oyd} 
  R.~Alonso, B.~Grinstein, and J.~Martin Camalich,
  Phys.\ Rev.\ Lett.\  {\bf 118}, no. 8, 081802 (2017)
  [arXiv:1611.06676 [hep-ph]].
  
\bibitem{Colquhoun:2015oha} 
  B.~Colquhoun {\it et al.} [HPQCD Collaboration],
  Phys.\ Rev.\ D {\bf 91}, no. 11, 114509 (2015)
  [arXiv:1503.05762 [hep-lat]].
  
        \bibitem{Belyaev:2012qa} 
  A.~Belyaev, N.~D.~Christensen, and A.~Pukhov,
  Comput.\ Phys.\ Commun.\  {\bf 184}, 1729 (2013)
  [arXiv:1207.6082 [hep-ph]].
  

  
  
\bibitem{Belyaev:2005ew} 
  A.~Belyaev, C.~Leroy, R.~Mehdiyev, and A.~Pukhov,
  JHEP {\bf 0509}, 005 (2005)
  [hep-ph/0502067].

\bibitem{Nadolsky:2008zw} 
  P.~M.~Nadolsky, H.~L.~Lai, Q.~H.~Cao, J.~Huston, J.~Pumplin, D.~Stump, W.~K.~Tung, and C.-P.~Yuan,
  Phys.\ Rev.\ D {\bf 78}, 013004 (2008)
  [arXiv:0802.0007 [hep-ph]].

\bibitem{Khachatryan:2015qda} 
  V.~Khachatryan {\it et al.} [CMS Collaboration],
  Phys.\ Rev.\ D {\bf 93}, no. 3, 032005 (2016)
  Erratum: [Phys.\ Rev.\ D {\bf 95}, no. 3, 039906 (2017)]
  [arXiv:1509.03750 [hep-ex]].

\end{thebibliography}
\end{document}